\documentclass[fleqn,usenatbib]{mnras}

\usepackage{newtxtext,newtxmath}

\usepackage[T1]{fontenc}

\DeclareRobustCommand{\VAN}[3]{#2}
\let\VANthebibliography\thebibliography
\def\thebibliography{\DeclareRobustCommand{\VAN}[3]{##3}\VANthebibliography}

\usepackage{graphicx}	
\usepackage{amsmath}	

\usepackage{natbib}
\usepackage{upgreek}

\def\be{\begin{equation}}
\def\ee{\end{equation}}
\newcommand\code[1]{\textsc{\MakeLowercase{#1}}}

\def\gsim{\lower.5ex\hbox{\gtsima}} 
\def\lsim{\lower.5ex\hbox{\ltsima}} 
\def\gtsima{$\; \buildrel > \over \sim \;$} 
\def\ltsima{$\; \buildrel < \over \sim \;$} \def\gsim{\lower.5ex\hbox{\gtsima}} 
\def\lsim{\lower.5ex\hbox{\ltsima}} 
\def\simgt{\lower.5ex\hbox{\gtsima}} 
\def\simlt{\lower.5ex\hbox{\ltsima}}

\def\mum{\mu {\rm m}}

\def\aco{\rm M_{\odot}\,(K\, km\, s^{-1}\, pc^2)^{-1}}

\def\Sg{$\Sigma_{\rm gas}$}
\def\S*{$\Sigma_{\rm SFR}$}
\def\Scii{$\Sigma_{\rm [CII]}$}

\def\Soiii{$\Sigma_{\rm [OIII]}$}

\def\OIII{\hbox{[O~$\scriptstyle\rm III $]}}
\def\CII{\hbox{[C~$\scriptstyle\rm II $]}}

\def\HII{\hbox{H$\,\scriptstyle\rm II$}}

\def\ks{\kappa_{\rm s}}

\defcitealias{vallini2020}{V20}
\defcitealias{vallini2021}{V21}
\defcitealias{ferrara2019}{F19}

\title[Resolved KS relation in the EoR]{Spatially resolved Kennicutt-Schmidt relation at $z\approx 7$ and its connection with the interstellar medium properties}

\author[Vallini et al.]{Livia Vallini$^{1,2}$,\thanks{E-mail: livia.vallini@inaf.it (LV)}
Joris Witstok$^{3,4}$,
Laura Sommovigo$^{2}$,
Andrea Pallottini$^{2}$,
Andrea Ferrara$^{2}$,
\newauthor
Stefano Carniani$^{2}$,
Mahsa Kohandel$^{2}$,
Renske Smit$^{5}$,
Simona Gallerani$^{2}$,
Carlotta Gruppioni$^{1}$
\\
$^{1}$INAF-Osservatorio di Astrofisica e Scienza dello Spazio, via Gobetti 93/3, I-40129, Bologna, Italy\\
$^{2}$Scuola Normale Superiore, Piazza dei Cavalieri 7, I-56126, Pisa, Italy\\
$^{3}$Kavli Institute for Cosmology, University of Cambridge, Madingley Road, Cambridge CB3 0HA, UK\\
$^{4}$Cavendish Laboratory, University of Cambridge, 19 JJ Thomson Avenue, Cambridge CB3 0HE, UK\\
$^{5}$Astrophysics Research Institute, Liverpool John Moores University, 146 Brownlow Hill, Liverpool L3 5RF, UK
}
\date{Accepted XXX. Received YYY; in original form ZZZ}

\pubyear{2023}

\begin{document}
\label{firstpage}
\pagerange{\pageref{firstpage}--\pageref{lastpage}}
\maketitle

\begin{abstract}
We exploit moderately resolved \hbox{[O~$\scriptstyle\rm III $]}, \hbox{[C~$\scriptstyle\rm II$]} and dust continuum ALMA observations to derive the gas density ($n$), the gas-phase metallicity ($Z$) and the deviation from the Kennicutt-Schmidt (KS) relation ($\kappa_s$) on $\approx \, \rm sub-kpc$ scales in the interstellar medium (ISM) of five bright Lyman Break Galaxies at the Epoch of Reionization ($z\approx 7$). To do so, we use \texttt{\texttt{GLAM}}, a state-of-art, physically motivated Bayesian model that links the \CII~and \OIII~surface brightness ($\Sigma_{\rm [CII]}$, $\Sigma_{\rm [OIII]}$) and the SFR surface density ($\Sigma_{\rm SFR}$) to $n$, $\kappa_s$, and $Z$. All five sources are characterized by a central starbursting region, where the $\Sigma_{\rm gas}$ vs $\Sigma_{\rm SFR}$ align $\approx$10$\times$ above the KS relation ($\kappa_s\approx10$). This translates into gas depletion times in the range $t_{\rm dep}\approx 80-250$ Myr. The inner starbursting centers are characterized by higher gas density ($\log (n/{\rm cm^{-3}}) \approx 2.5-3.0$) and higher metallicity ($\log (Z/Z_{\odot}) \approx -0.5$) than the galaxy outskirts. We derive marginally negative radial metallicity gradients  ($\nabla \log Z \approx -0.03 \pm 0.07$ dex/kpc), and a dust temperature ($T_d\approx32-38$ K) that anticorrelates with the gas depletion time.

\end{abstract}




\section{Introduction} \label{sec:intro}
The Epoch of Reionization (EoR) represents a critical phase of the Universe evolution, and its study is one of the frontiers in modern astrophysics \citep[e.g.][]{robertson2022}. During the EoR, the first galaxies started to rapidly form stars, which in turn began producing photons able to ionize the
surrounding gas – first the interstellar medium (ISM), and eventually the intergalactic medium \citep[][for a review]{dayal2018}. For this reason, shedding light on how the gas is converted into stars \citep[e.g.][]{tacconi2020}, and how this process is influenced by the ISM properties holds the key to understanding the evolution of cosmic reionization.

At low and intermediate redshifts, the so-called Kennicutt-Schmidt (KS) relation\footnote{The star formation rate (gas) surface density is expressed in units of $\rm M_{\odot} \, yr^{-1} \, kpc^{-2}$ ($\rm M_{\odot} \, kpc^{-2}$).} $\Sigma_{\rm SFR}\approx 10^{-12}\kappa_s \Sigma_{\rm gas}^{1.4}$, linking the star formation rate (SFR) and the gas surface densities (\S*, \Sg~respectively) is well established \citep{schmidt1959, kennicutt1998, heiderman10, delosreyes2019}. The ``burstiness" parameter, $\kappa_s$, was first introduced in \citet{ferrara2019} to quantify the deviation from the KS relation that might occur in the high-$z$ Universe. Galaxies with $\kappa_s > 1$ show a larger SFR per unit area with respect to those located on the KS relation, i.e. they tend to be starburst. At high redshifts, values in the range $\kappa_s = 10-100$ have been measured for massive/rare sub-millimeter galaxies for which spatially-resolved data of cold gas tracers, namely low-$J$ CO lines, are available \citep[e.g.][]{hodge2015, chen2017}. Spatially-resolved low-$J$ CO detections in galaxies representative of the bulk population in the EoR are instead time demanding even with state-of-art radio/sub-mm facilities, unless taking advantage of strong gravitational lensing \citep[][at z$\approx1$]{nagy2023}. This is due to the efficient CO photodissociation at low metallicity and dust abundance \citep{bolatto2013, wolfire2022}, and because of the increasing temperature of the CMB \citep[][]{dacunha2013, vallini2015} against which the lines are observed. Only a few mid-$J$ CO detections have been reported so far \citep[e.g.][]{pavesi2019, ono2022} but none of them are spatially resolved even by the Atacama Large Millimeter/submillimeter Array (ALMA), thus hampering the measure of the size of the emitting area and ultimately the derivation of \Sg.\\

In recent years an alternative, indirect, method has been proposed to infer the location of EoR sources with respect to the KS relation. This is done by linking their $\kappa_s$ to the relative surface brightness ratios of bright neutral (e.g. \CII~158$\mu$m) versus ionized (e.g. \OIII~88$\mu$m, CIII]$\lambda$1907,1909 doublet) gas tracers \citep{ferrara2019, vallini2020, vallini2021, markov2022} that can be spatially resolved by ALMA \citep[][]{herrera-camus2022, akins2022, molyneux2022, witstok2022, posses2023} and JWST \citep[e.g.][]{hsiao2023}. 
A starburst source has a larger ionization parameter, $U$, producing a  correspondingly larger ionized gas co\-lumn density, as compared to a galaxy with the same \Sg~but lying on the KS relation \citep{ferrara2019}. These conditions boost (quench) ionized (neutral) gas tracers and, together with the gas density ($n$) and metallicity ($Z$), concur in determining the surface brightness ratios \citep{kohandel2023}.

By leveraging this method, \citet{vallini2021} analy\-zed the nine EoR Lyman Break Galaxies (LBGs) that had joint (albeit only barely resolved) [C~$\scriptstyle\rm II $]-\OIII~detections at the time \citep[][]{laporte2017, tamura2019,harikane2020, bakx2020, carniani2020}, obtaining $\kappa_s = 10-100$. These high burstiness parameters, in agreement with expectations from cosmological zoom-in simulations \citep{pallottini:2019,pallottini2022}, suggest ISM conditions favouring an efficient conversion of gas into stars (short depletion times), with starburst episodes producing bright \OIII~emission from \HII~regions \citep[e.g.][]{cormier2019, harikane2020}. Also, the gas metallicity and density were found to be relatively high ($Z=0.2-0.5\, Z_\odot$, and $n =10^{2-3} \,\rm{cm^{-3}}$, respectively) in agreement with independent analysis carried out on the same objects \citep{jones2020, yang2020}. 

A recent study of three $z\approx7$ galaxies from REBELS \citep{bouwens2022} supports the tight relation between galaxy burstiness and \OIII/\CII~ratios. In this case, low \OIII/\CII~have been explained with the weak ionizing field resulting from the non-starbursting nature of the sources \citep{algera2023}. 
The lack of recent bursts is also likely the cause \citep[e.g.][]{sommovigo2020} of their cold dust temperatures. The sources analyzed by \citet{algera2023} seem, however, to be an outlier with respect to the average conditions of EoR galaxies with below-average [OIII]$\lambda \lambda 4959,\, 5007+$H$\beta$ equivalent widths compared to the known high-$z$ population.

{\color{black} From the theoretical side, an increasing number of simulations and models developed to interpret \CII~\citep[e.g.][]{vallini2015, lagache2018, pallottini2022}, \OIII~\citep[e.g.][]{moriwaki2018, arata2020, katz2022} and dust continuum emission \citep[e.g.][]{behrens2018, dicesare2023}, find high turbulence \citep[e.g.][]{kohandel2020}, strong radiation fields \citep[e.g.][]{katz2022}, high densities, and warm dust temperatures \citep[e.g.][]{sommovigo2021}, to be common on sub-kpc scales in the ISM of star forming galaxies in the EoR.}\\

The goal of this work is to push further the study of the link between the KS relation, and ISM/dust properties in the first galaxies by leveraging the spatially resolved \CII, \OIII, and dust continuum data recently presented by \citet{witstok2022} in a sample of five bright LBGs at $z\approx7$. Our aim is to investigate the sub-kpc relation between the burstiness parameter, gas density and metallicity and study their connection with global values that can be inferred from unresolved data.

The paper is structured as follows: in Sec. \ref{sec:data} we summarize the sample and data used in this analysis, in Sec. \ref{sec:model} we illustrate the model. The results are outlined in Sec. \ref{sec:results} while we discuss the implications and present our conclusions in the final Section \ref{sec:conclusions}.

\section{Sample and data} \label{sec:data}
Details regarding the sample and data reduction can be found in \citet{witstok2022}, however we summarize the key points here. We considered all available ALMA data sets of \CII~158$\upmu$m (2015.1.01111.S, 2017.1.00604.S, 2019.1.01611.S, PI: Smit, 2018.1.00085.S, PI: Schouws, 2018.1.01359.S, PI: Aravena, 2015.1.00540.S, 2018.1.00933.S, PI: Bowler) and \OIII~88 $\upmu$m (2018.1.00429.S, 2019.1.01524.S, PI: Smit) for the sample of LBGs at $z \sim 7$: COS-3018555981 (COS-3018, hereafter), COS-2987030247 (COS-2987, hereafter), UVISTA-Z-001, UVISTA-Z-007, and UVISTA-Z-019.
Data were calibrated and reduced with the automated pipeline of the Common Astronomy Software Application \citep[\textsc{casa};][]{mcmullin2007}. In cases where the continuum is robustly detected (i.e. next to the \CII~emission in COS-3018, UVISTA-Z-001, and UVISTA-Z-019 and next to the \OIII~emission in UVISTA-Z-001), we first performed continuum subtraction using the \textsc{uvcontsub} task in \textsc{casa}. After this step, we created images with the \textsc{tclean} task both under natural and several Briggs weightings. We tuned the weighting and/or taper scheme to match the beam sizes as closely as possible, using natural weighting (and a small taper, if required) for the line observed with highest spatial resolution and Briggs weighting for the other. The robust parameter has been tuned to the highest resolution achievable while maintaining a reasonable signal-to-noise ratio. The resulting matched beam sizes ($\theta \approx 0.4''-0.5''$) for \CII~and \OIII~are listed in Table~2 of \citet{witstok2022}. Finally, we regridded images of the \OIII~and \CII, obtained by integrating along the frequency axis over the full width at half maximum (FWHM) of the line, to a common coordinate mesh with the \textsc{reproject} package in \textsc{astropy}.

\begin{figure*}
    \centering
    \includegraphics[scale=0.4]{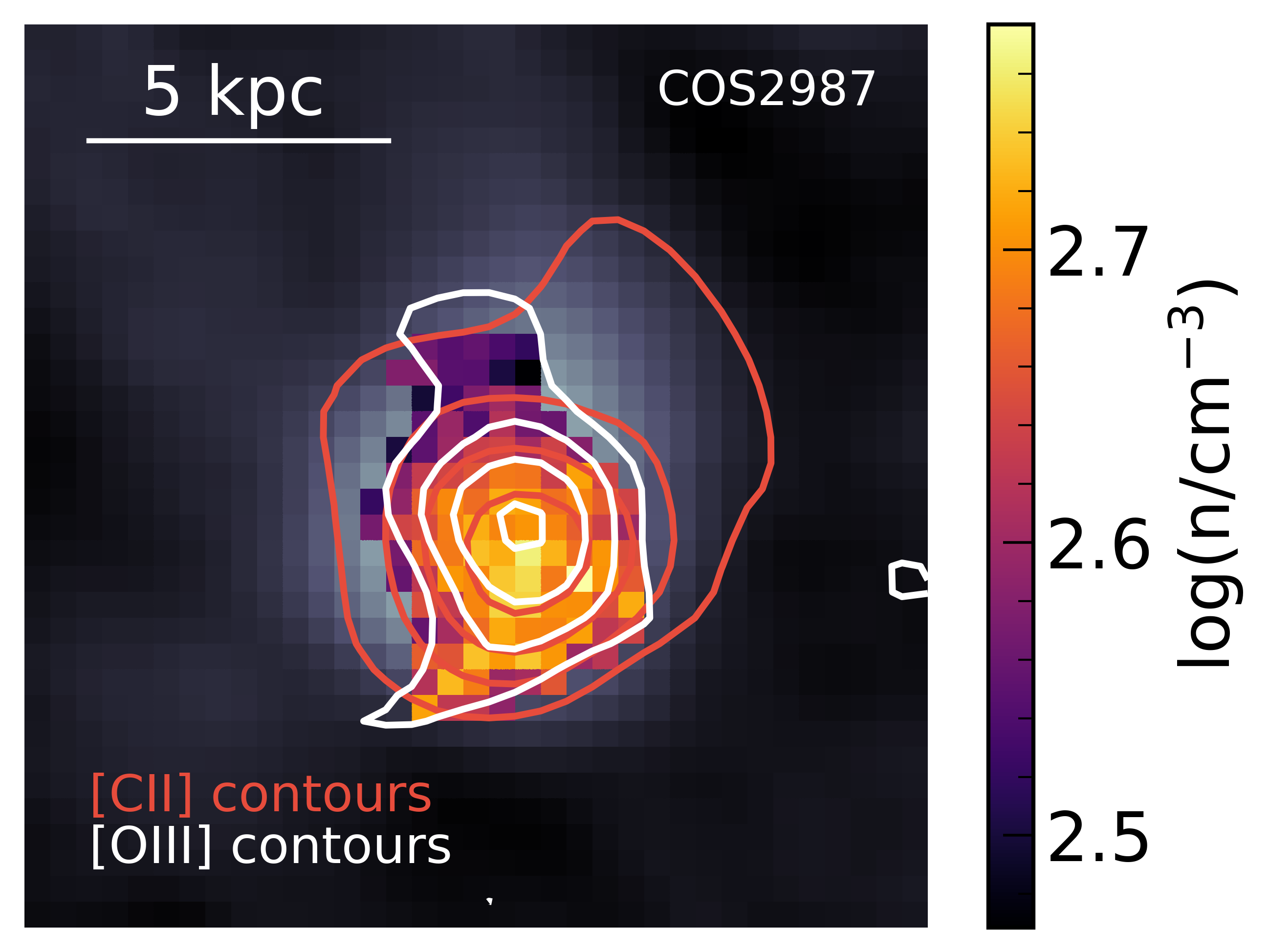}
    \includegraphics[scale=0.4]{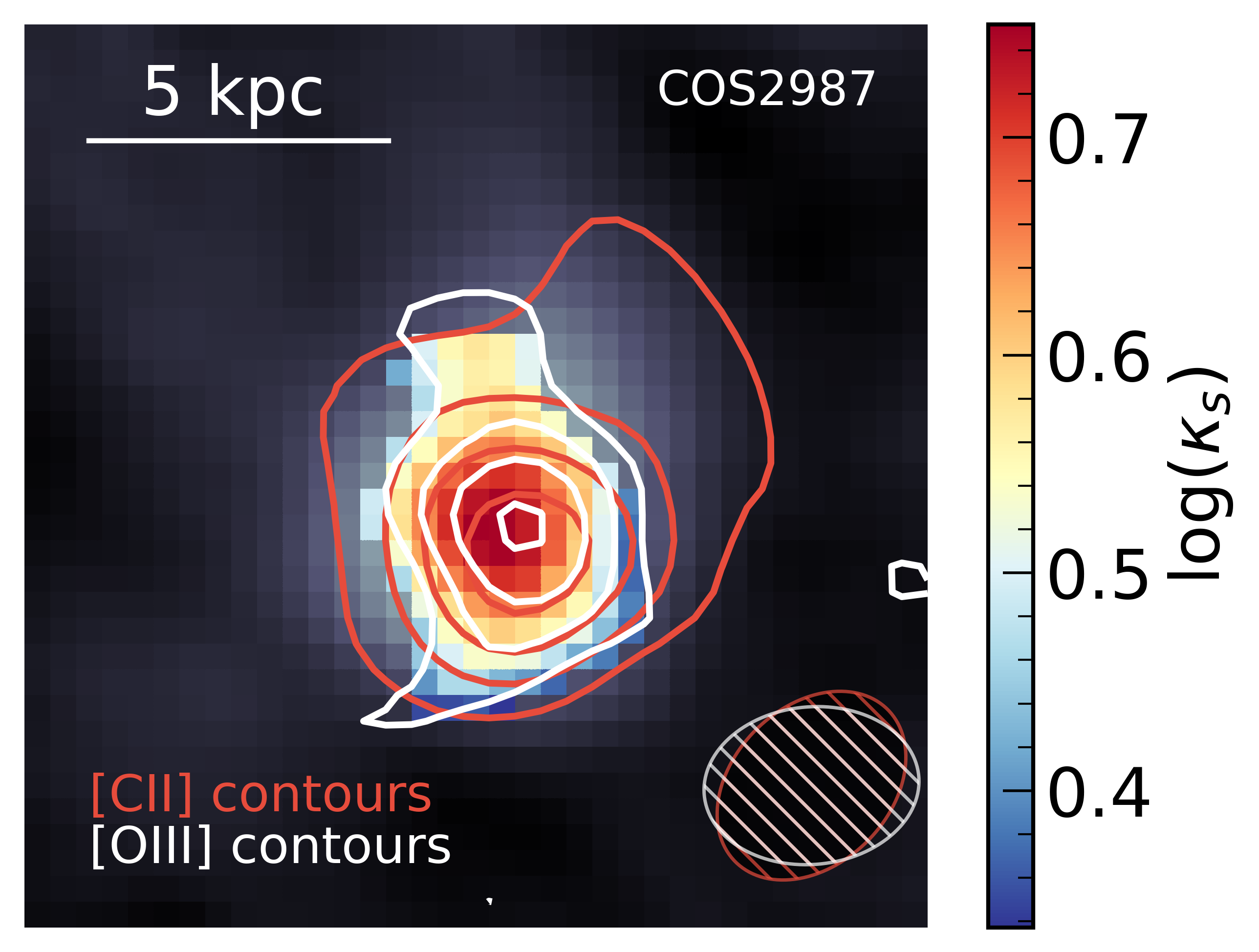}
    \includegraphics[scale=0.4]{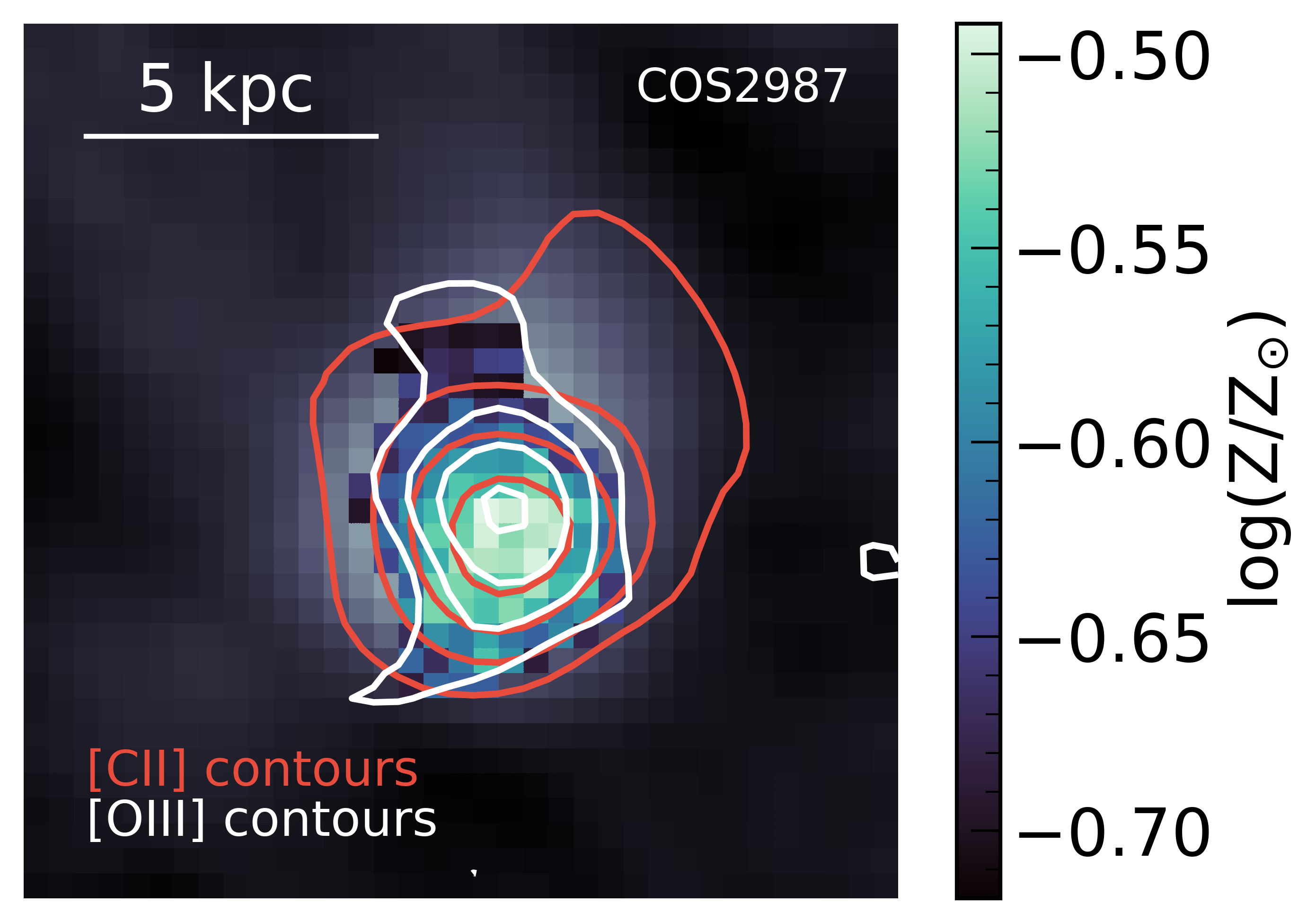}
    \includegraphics[scale=0.4]{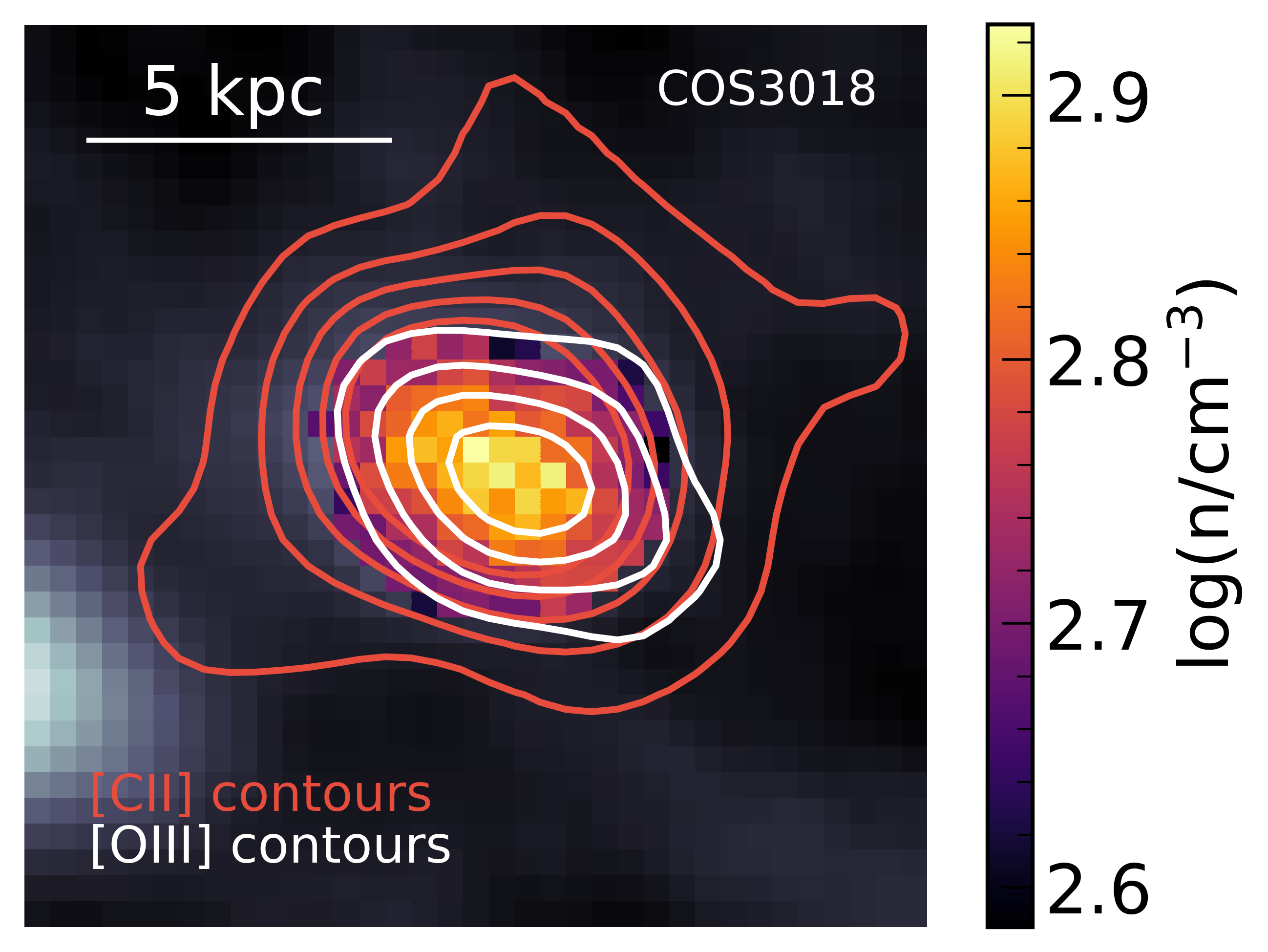}
    \includegraphics[scale=0.4]{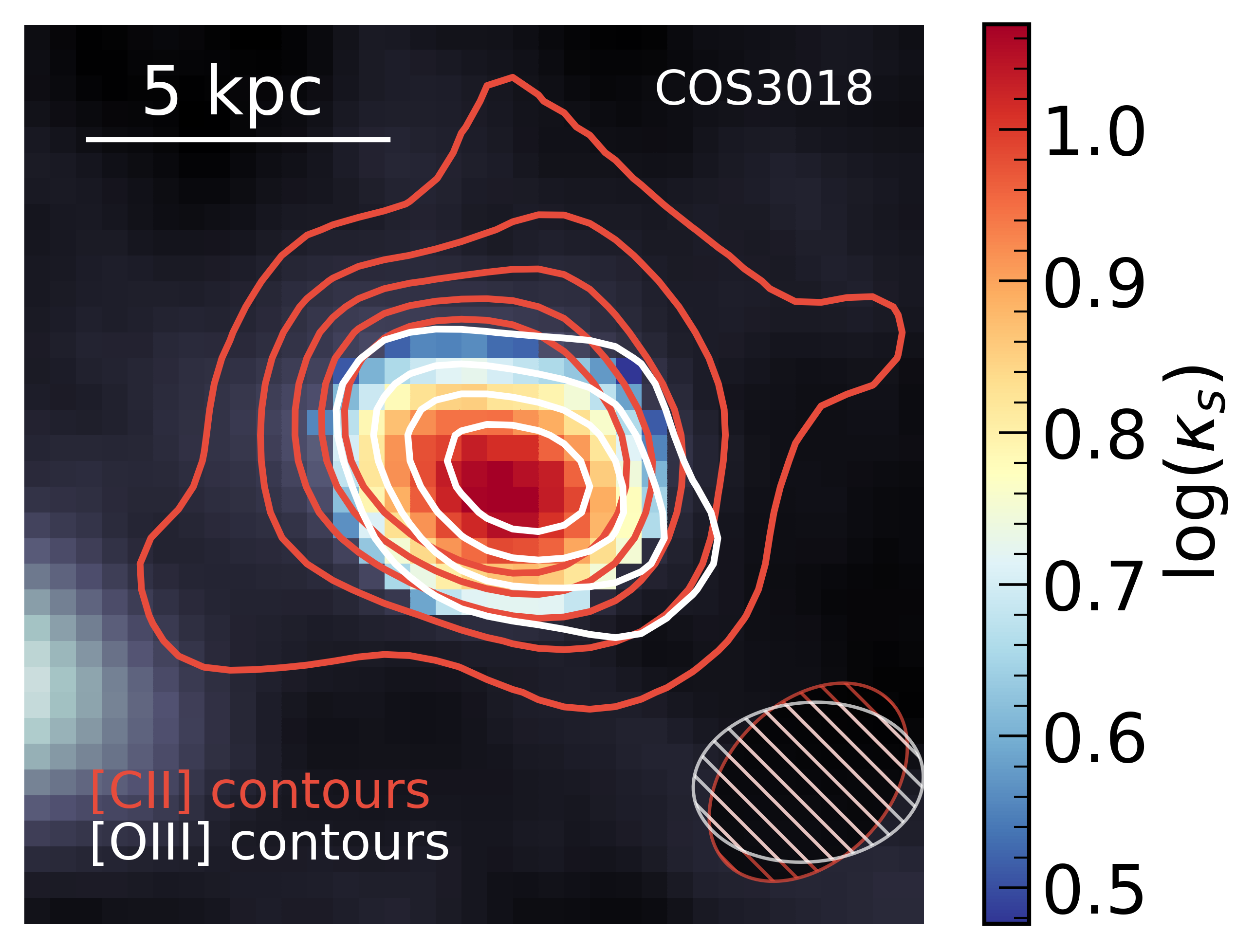}
    \includegraphics[scale=0.4]{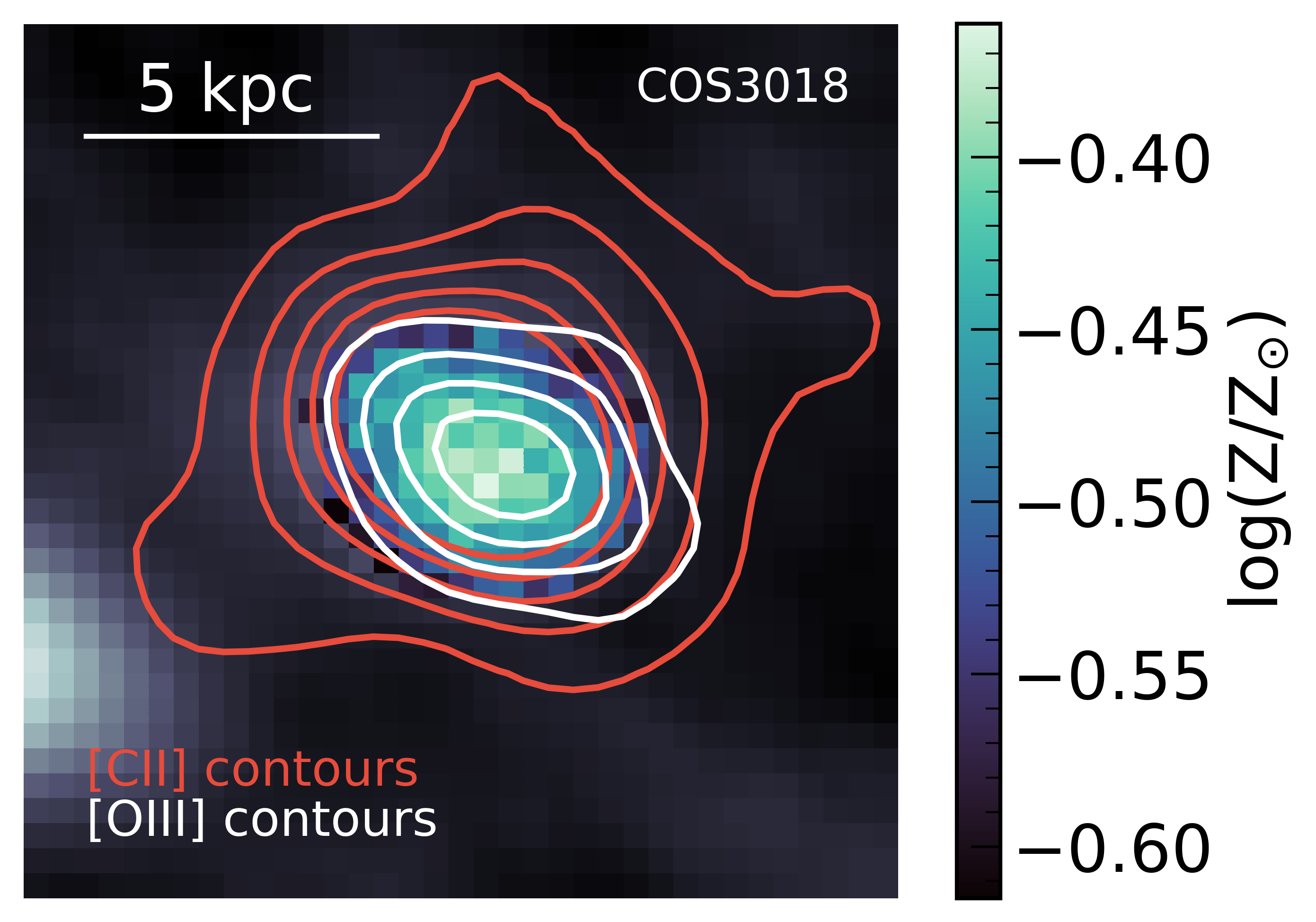}
    \includegraphics[scale=0.4]{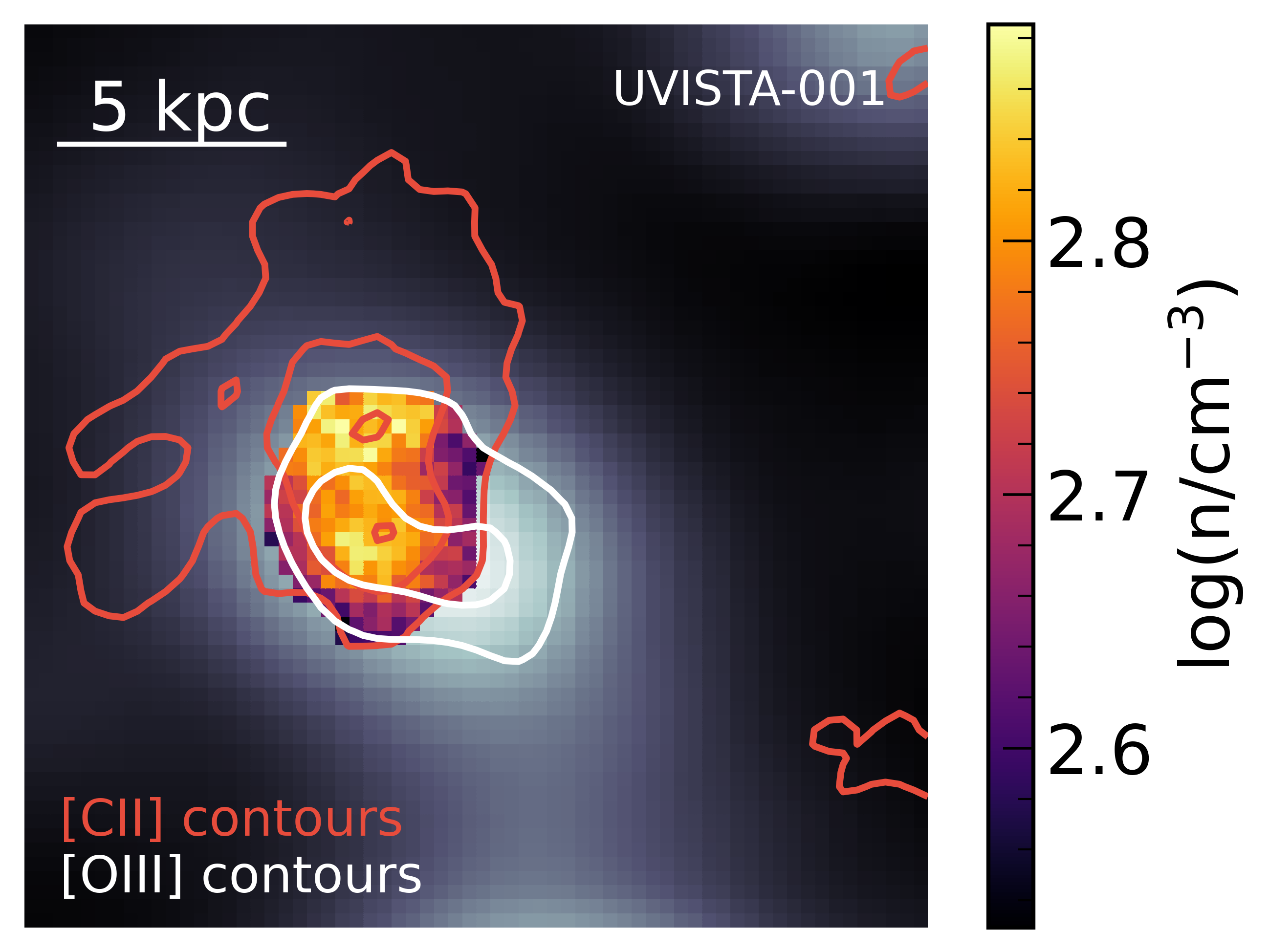}
    \includegraphics[scale=0.4]{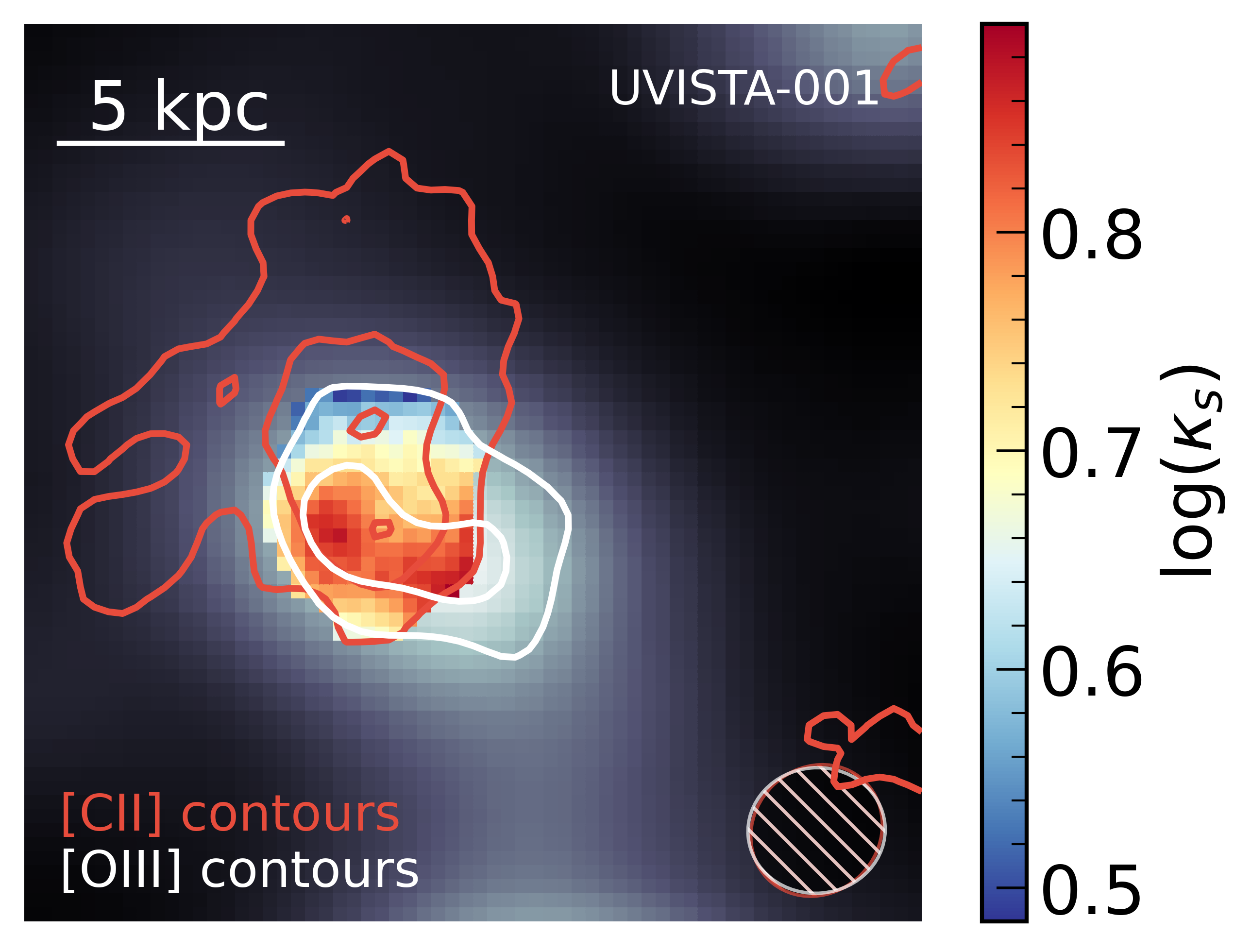}
    \includegraphics[scale=0.4]{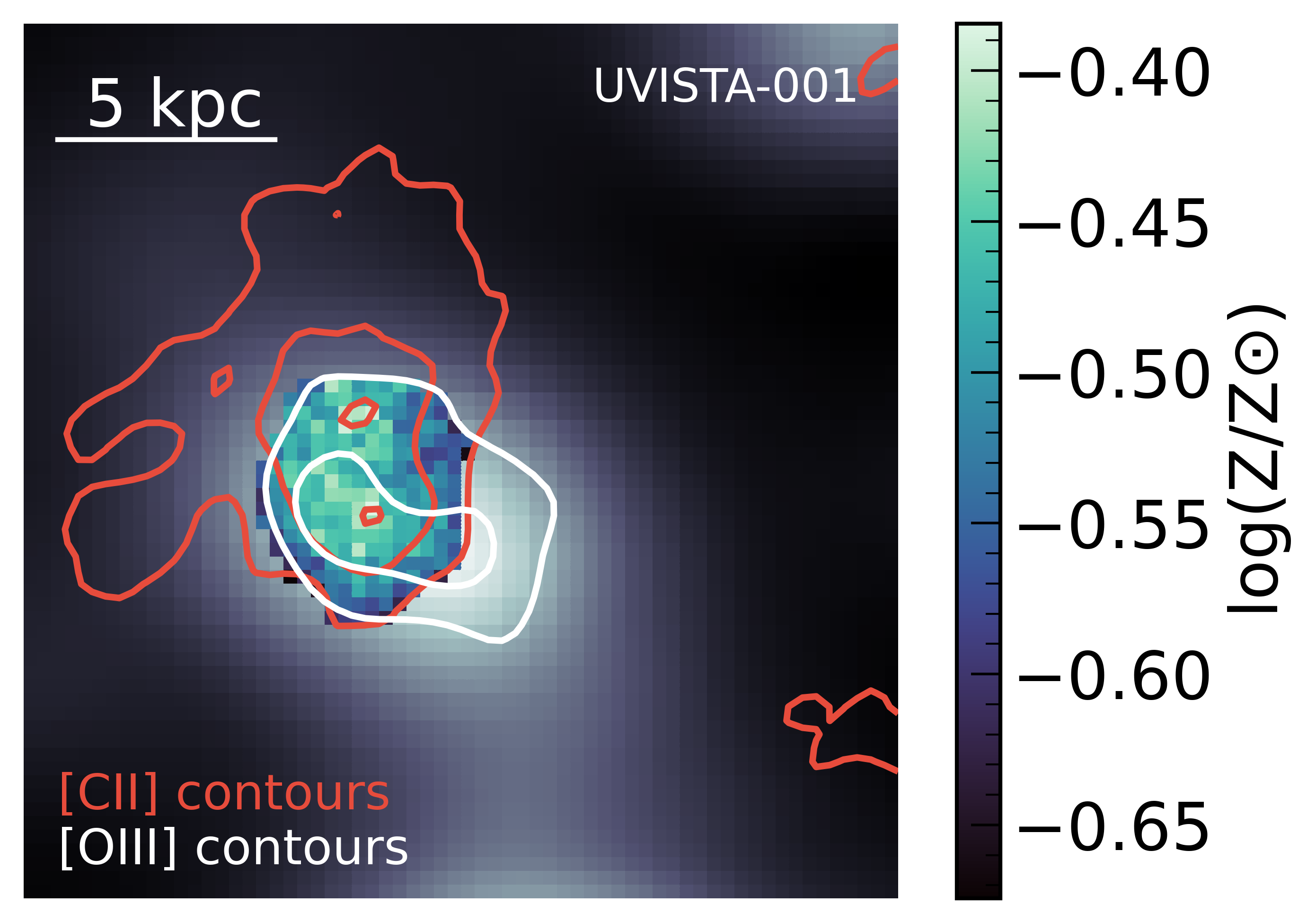}
    \includegraphics[scale=0.4]{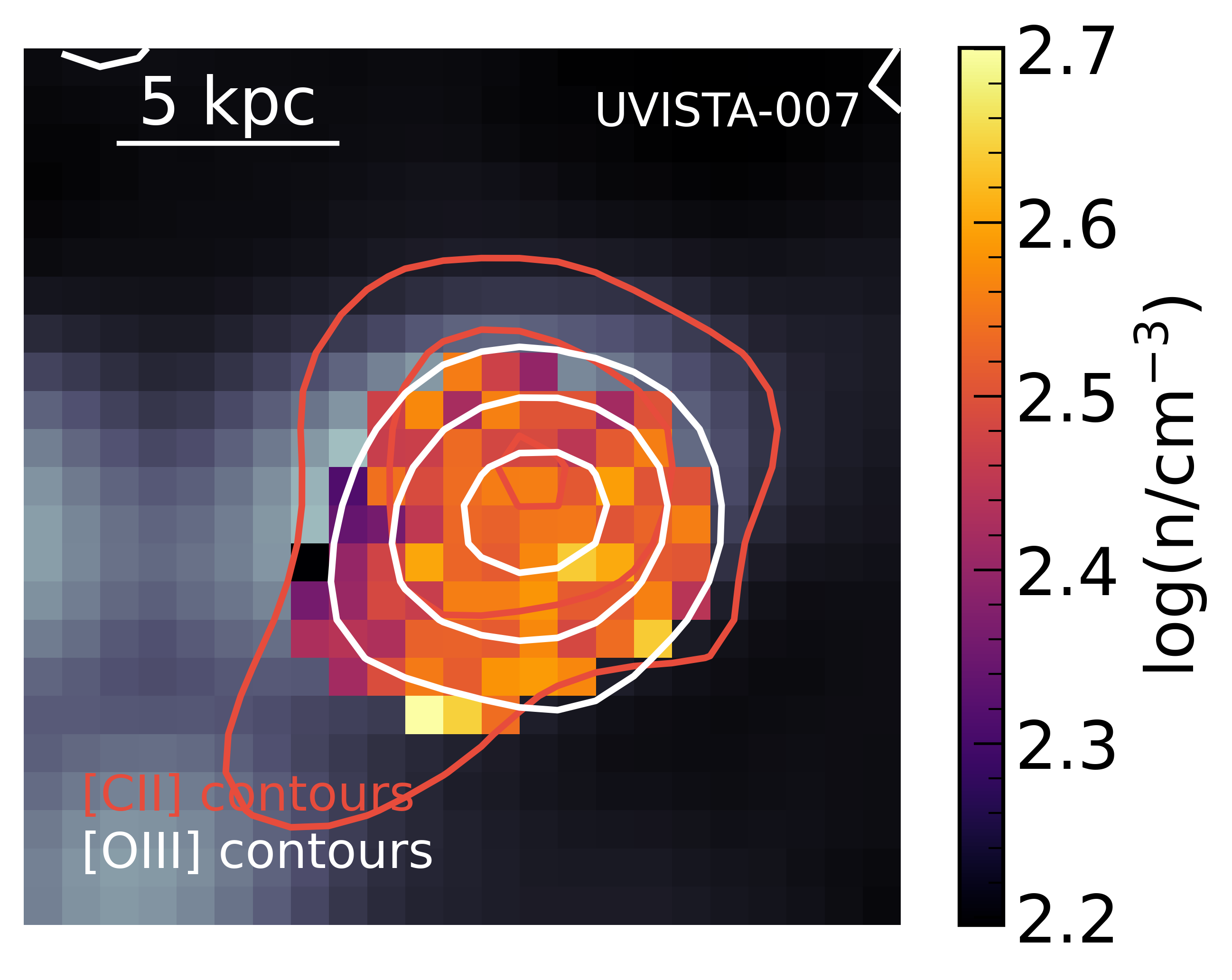}
    \includegraphics[scale=0.4]{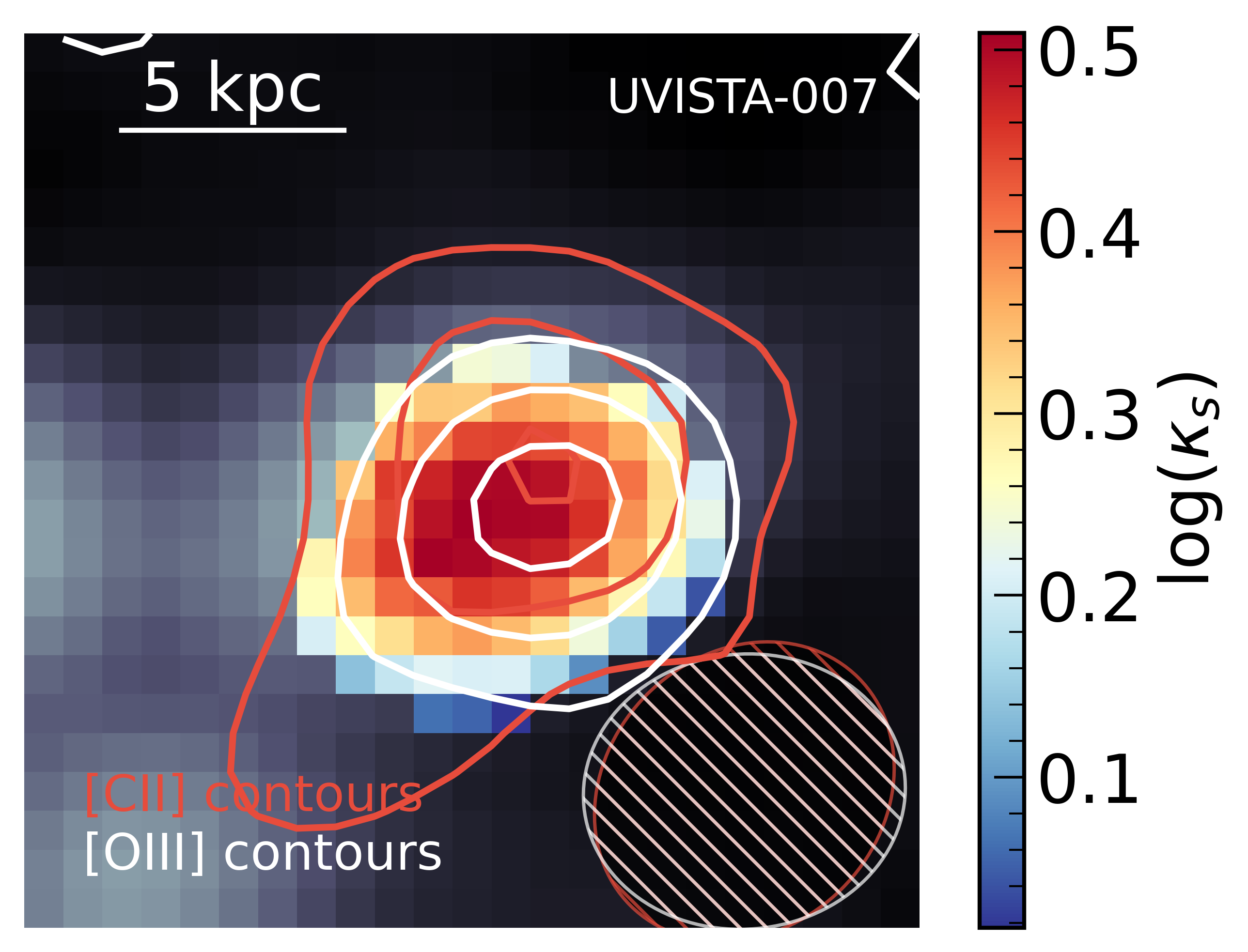}
    \includegraphics[scale=0.4]{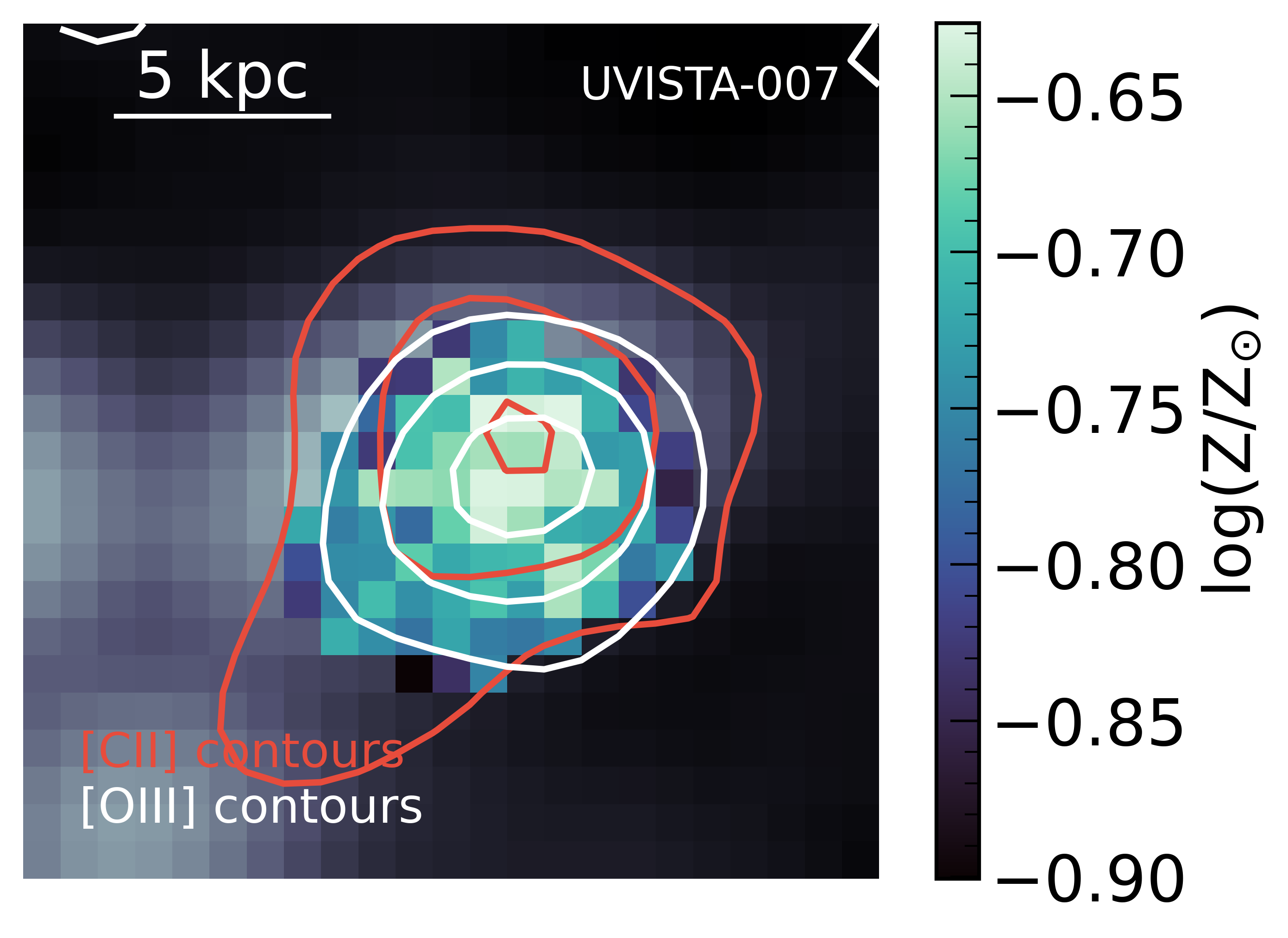}
    \includegraphics[scale=0.4]{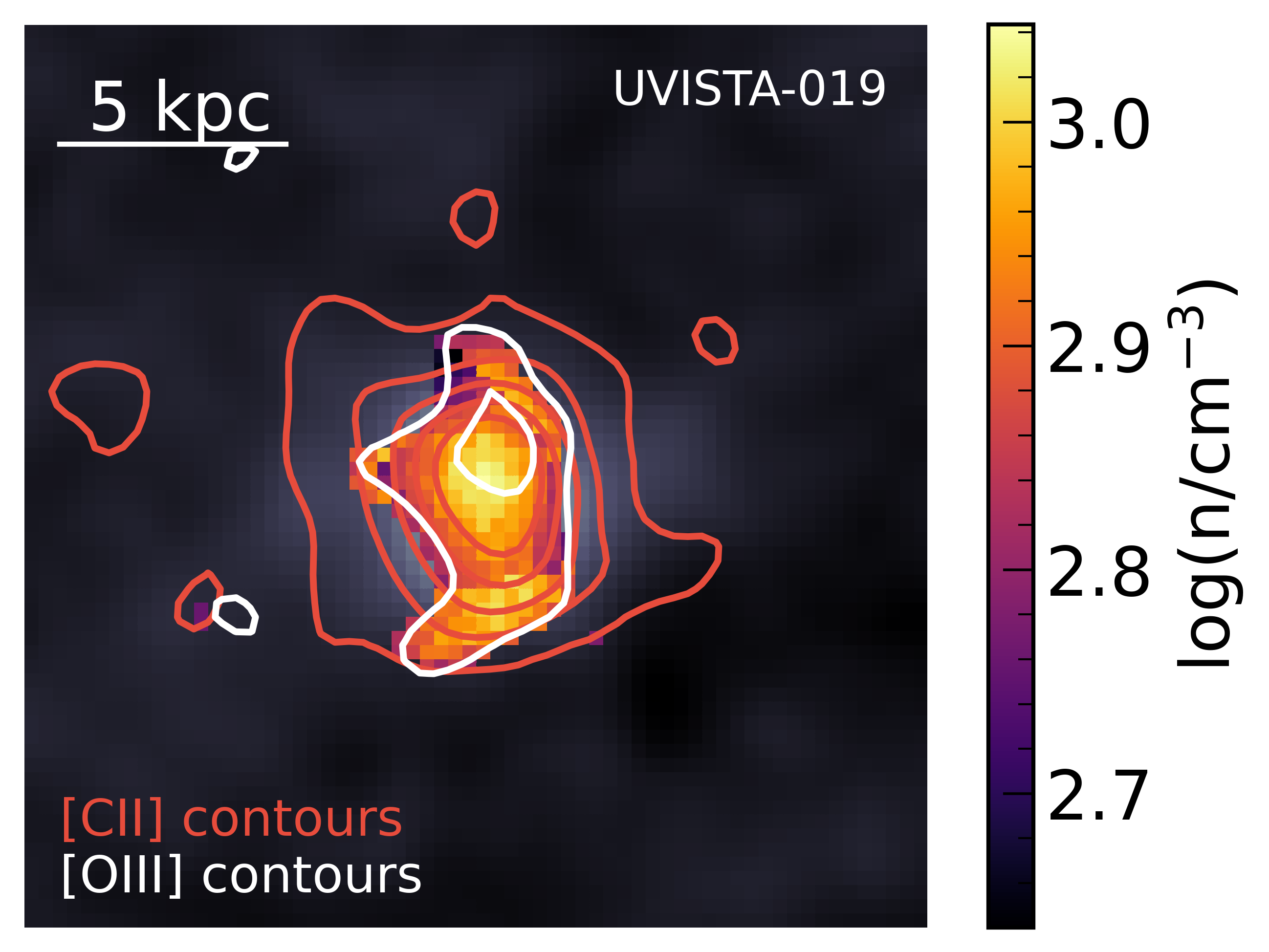}
    \includegraphics[scale=0.4]{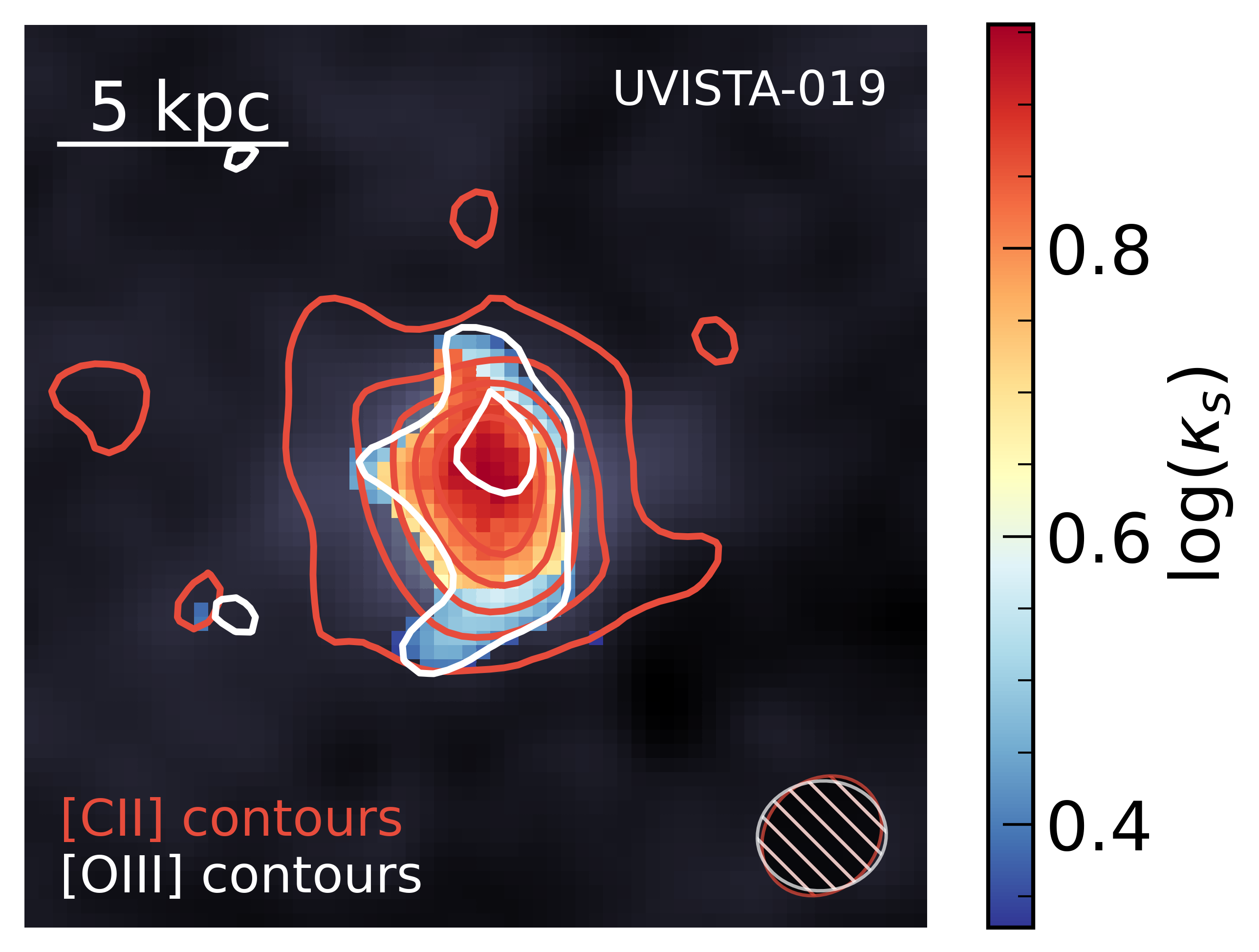}
    \includegraphics[scale=0.4]{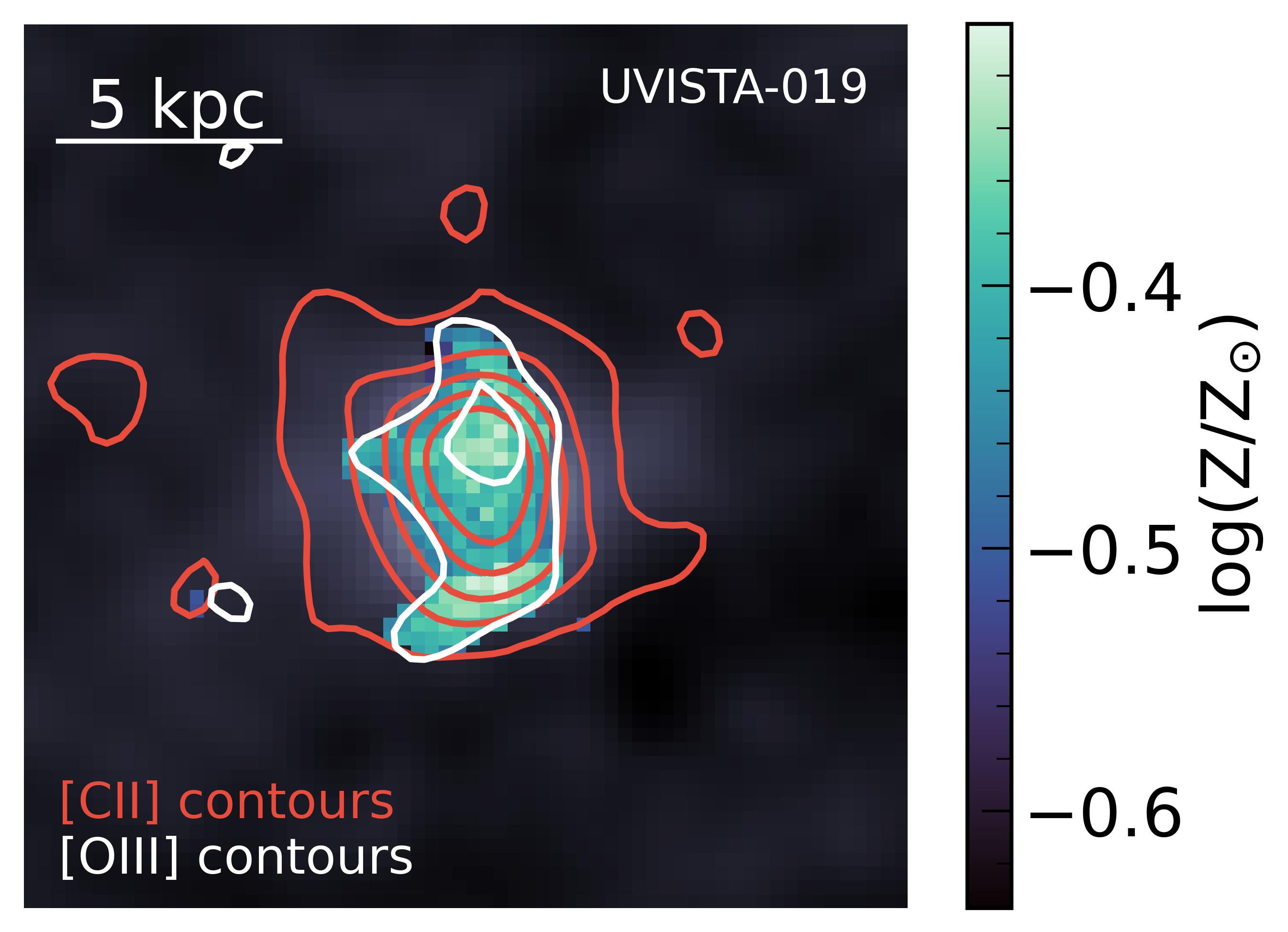}

    \caption{The derived gas density ($n$, left column), deviation from the KS relation ($\kappa_s$, central column), and gas metallicity ($Z$, right column) for the five galaxies (COS-2987, COS-3018, UVISTA-Z-001, UVISTA-Z-007, UVISTA-Z-019, from top to bottom) analyzed in this work with \texttt{GLAM}. The \CII~and \OIII~contours (red and white, respectively) are overplotted onto the HST rest-frame UV images (in background). All the contours start at $3\sigma$. The matched \CII~and \OIII~beam sizes used for the present analysis are indicated in the bottom right corner of the $\kappa_s$ maps (central column).}
    \label{fig:GLAM_maps}
\end{figure*}
Both UV and IR luminosity maps were regridded to the same pixel grid as the \OIII~and \CII~lines. We used the same imaging parameters for the dust continuum at rest-frame wavelength $\lambda_{\text{emit}} \sim 160 \, \mathrm{\upmu m}$ (since band-6 measurements had the most significant detections) to achieve a beam nearly identical to that of the \CII~line. The IR luminosity ($L_{\rm IR}$, $8-1000\,\rm \mu m$) was calculated using the global best-fit spectral energy distribution (SED, see Section~4 in \citealt{witstok2022} and Table 3 for the best-fit dust temperatures, $T_d\approx 29-60\, \rm K$) as a template for rescaling the $\sim 160 \, \mathrm{\upmu m}$ flux in each pixel.
The UV continuum was convolved with an effective \citet{richardson1972, lucy1974} beam to match the point spread function (PSF) of the dust-continuum emission. 

\section{Model} 
\label{sec:model}
The derivation and study of the ISM properties presented in this paper is based on \texttt{\texttt{GLAM}}\footnote{\texttt{GLAM}: Galaxy Line Analyzer with MCMC, is publicly available at \url{https://lvallini.github.io/MCMC_galaxyline_analyzer/} }  \citep[][hereafter \citetalias{vallini2020}, \citetalias{vallini2021}]{vallini2020, vallini2021}. \texttt{GLAM} is a tool to perform Bayesian inference that is based on a physically motivated model for the analytical treatment of the radiative transfer of ionizing ($ h\nu>13.6 \rm \, eV$, EUV) and non-ionizing ($6 {\rm \, eV}< h\nu < 13.6 {\rm \, eV}$, FUV) photons in the ISM of galaxies \citep[][hereafter \citetalias{ferrara2019}]{ferrara2019}.

\citetalias{ferrara2019} enables the computation of the surface brightness of lines excited either in the ionized and/or in the photodissociation region \citep[PDR;][]{hollenbach1999, wolfire2022} of a gas slab illuminated by ultraviolet (UV) radiation from newly formed stars. The surface brightness of the lines is determined by the average gas density ($n$) of the \HII/PDR environment -- {\color{black} characterized by electron density $n_e$ and neutral gas density $n_H$, respectively}\footnote{{\color{black}Both $n_e$ and $n_H$ can be expressed as a function of $n$. In the ionized layer, $n_e=x_e n\approx n$  assuming an ionized fraction $x_e\approx 1$, while in the PDR the neutral gas density $n_H=(1-x_e) n \approx n$ given that $x_e\approx 0$.}} --, the dust-to-gas ratio, ($\mathcal{D}\propto Z$, where $Z$ is the gas metallicity), and ionisation parameter, $U$. The latter, can be expressed in terms of observed quantities by
de\-ri\-vi\-ng its relation ($U \propto$ \S*/\Sg, see eqs. 38 and 40 in \citetalias{ferrara2019}) with the star formation rate surface density (\S*) and the gas surface density (\Sg), which in turn are connected through the star formation law. This leaves us with the $\kappa_s$ parameter, describing the burstiness of the galaxy. \texttt{\texttt{GLAM}} adopts a Markov Chain Monte Carlo (MCMC) algorithm \citep[implemented with \texttt{emcee},][]{foreman2013} to search for the posterior probability of the model parameters ($n$, $\ks$, $Z$) that reproduce the observed \CII~surface brightness ($\Sigma_{\rm [CII]}$), \OIII~surface brightness ($\Sigma_{\rm [OIII]}$), and the SFR surface density ($\Sigma_{\rm SFR}$). \texttt{\texttt{GLAM}} accounts for the observed errors ($\delta_{\rm[CII]}$, $\delta_{\rm[OIII]}$, $\delta_{\rm SFR}$) and can accept also different lines (e.g. CIII]$\lambda$1909 instead of \OIII) as input \citep[\citetalias{vallini2020},][]{markov2022}.

{\color{black} It is worth noting that the \citetalias{ferrara2019} model assumes a fixed O/C ratio \citep{asplund2009}, with carbon and oxygen abundances linearly scaling with metallicity, and a constant gas temperature in the ionized layer ($T=10000 \, \rm K$) and PDR ($T=100 \rm \, K$). The impact of the latter assumption has been tested against numerical radiative transfer calculations performed with CLOUDY \citep{ferland2017} over a wide range of ionization parameters and metallicities. Overall, the agreement is very good (see Figure 3 in \citetalias{ferrara2019}), with CLOUDY confirming both the amplitude and linear slope of the increasing \CII~flux with gas metallicity, along with the saturation of the \CII~flux for increasing $U$. In spite of the inevitable simplifications of an analytical model such as that of \citetalias{ferrara2019}, the differences with CLOUDY are relatively small (e.g. the \CII~flux is overestimated by \citetalias{ferrara2019} at most by $\approx 2$ at any $Z$). Moreover, as outlined in \citetalias{vallini2021}, the temperature in the ionized layer does not have a strong impact on the predicted \OIII~flux. The 88$\mu m$ line, and the other transition in the doublet (\OIII~at 52$\mum$). have similar excitation energy ($T_{\rm ex,88}\approx160 \rm \, K$ and $T_{\rm ex,52}\approx260 \rm \, K$) but different critical densities, hence for $T>1000$ K their ratio is only affected by the gas density \citep{palay2012}.}\\

Up to now, the exploitation of \texttt{\texttt{GLAM}} in hign-$z$ galaxies has been limited to the derivation of the galaxy-averaged $(n, \ks, Z)$ values \citep[\citetalias{vallini2020}, \citetalias{vallini2021},][]{markov2022} because of the relatively low ($>$ kpc scales) spatial resolution of the \CII, \OIII, and CIII] data \citep[][respectively]{smit2018, carniani2020, markov2022} used as input in the code.
The recent work by \citet{witstok2022}, which gathers moderate resolution ($\approx \rm \, kpc$) \CII~and \OIII~observations in a sample of $z\approx7$ galaxies, regridded to a common coordinate mesh of sub-kpc pixels, allows us for the first time the use of \texttt{\texttt{GLAM}} on a pixel-by-pixel basis, and the characterization at sub-kpc scales of the ISM properties of EoR sources.

We perform two types of analysis with \texttt{GLAM}. First, we fit a 2D Gaussian profile to the \CII, \OIII, UV, and IR continuum maps, to compute the global size of the emission ($r_{\rm [CII]}$,$r_{\rm [OIII]}$, $r_{\rm UV}$) and, with that, infer the mean \Scii$=L_{\rm [CII]}/\pi r_{\rm [CII]}^2$, \Soiii$=L_{\rm [OIII] }/\pi r_{\rm [OIII]}^2$, \S*$=({\rm SFR}_{\rm UV} + {\rm SFR}_{\rm IR})/\pi r_{\rm UV}^2$. The choice of a 2D gaussian profile, instead of e.g. an exponential one, is for consistency with the \citetalias{vallini2021} analysis $z\approx6-9$ sources with barely resolved observations. Second, we feed to the model the $\Sigma^{i}_{\rm [CII]}$, $\Sigma^{i}_{\rm [OIII]}$,$\Sigma^{i}_{\rm SFR}$ of each $i$-pixel of the grid (pixel size $\approx 0.3-0.8$ kpc, depending on the source), for which all the three quantities are above the $3\sigma$ level, to obtain spatially resolved derivation of the ISM parameters.

\section{Results}
\label{sec:results}
In this Section we presents our results and their implications, starting with an overview of the spatially resolved vs global gas density, metallicity, and burstiness values computed with \texttt{GLAM} (Sec. \ref{subsec:maps}). We then focus on key quantities from which we can infer insights on the ISM enrichment and baryon cycle  (Sec. \ref{subsec:radial_prof}), the conversion of the gas into stars (Sec. \ref{subsec:kennicutt_schmidt}), and the dust properties (Sec. \ref{subsec:dust}) in the EoR.
\subsection{Spatially resolved vs global ISM properties}
\label{subsec:maps}
In Figure \ref{fig:GLAM_maps}, we present the $n$, $\kappa_s$ and $Z$ maps, for the five galaxies in the \citet{witstok2022} sample, produced with \texttt{\texttt{GLAM}} by simultaneously fitting the $\Sigma^{i}_{\rm [CII]}$, $\Sigma^{i}_{\rm [OIII]}$, and the $\Sigma^{i}_{\rm SFR}$ in each pixel. We note that the central regions are characterized by higher gas density, burstiness parameter and metallicity, suggesting an inside-out star formation scenario. The sources are likely experiencing a burst in star formation in connection with the central \OIII~bright regions. The connection between recent bursts of star formation and high \OIII/\CII~ratios has been discussed on global galactic scales by several authors \citep{katz2017, arata2020, sugahara2022, pallottini2022, kohandel2023}, but this is the first time that we obtain a quantitative measure of spatially resolved trends in burstiness within EoR galaxies.

As outlined in Sec. \ref{sec:model}, for sources detected in the dust continuum (COS-3018, UVISTA-Z-001, and UVISTA-Z-019) we also considered the obscured star formation rate (SFR$_{\rm IR}$) when deriving the total $\Sigma_{\rm SFR}$. The SFR$_{\rm IR}$ is computed from the $L_{\rm IR}$ (see Sec \ref{sec:data}) using the conversion from \citet{kennicutt2012} in those pixels where the continuum is detected at $\geq 3\sigma$. 
For COS-3018 and UVISTA-Z-001 adding SFR$_{\rm IR}$ does not alter the smooth decreasing radial trends of ($n$, $\kappa_s$, $Z$) from the galaxy center towards the periphery, but in UVISTA-Z-019 the SFR$_{\rm IR}$ produces a sharp gradient in the $\kappa_s$ and $n$ values towards the center of the source. In the IR-detected central region both $\kappa_s$ and $n$ have higher values with respect to the neighbouring regions that are only UV detected. This finding can be even more prominent should the galaxy centres be characterized by warmer $T_d$ than the value derived from the global SED fitting procedure (see Sec. \ref{sec:data}).
\begin{figure}
    \centering
    \includegraphics[scale=0.5]{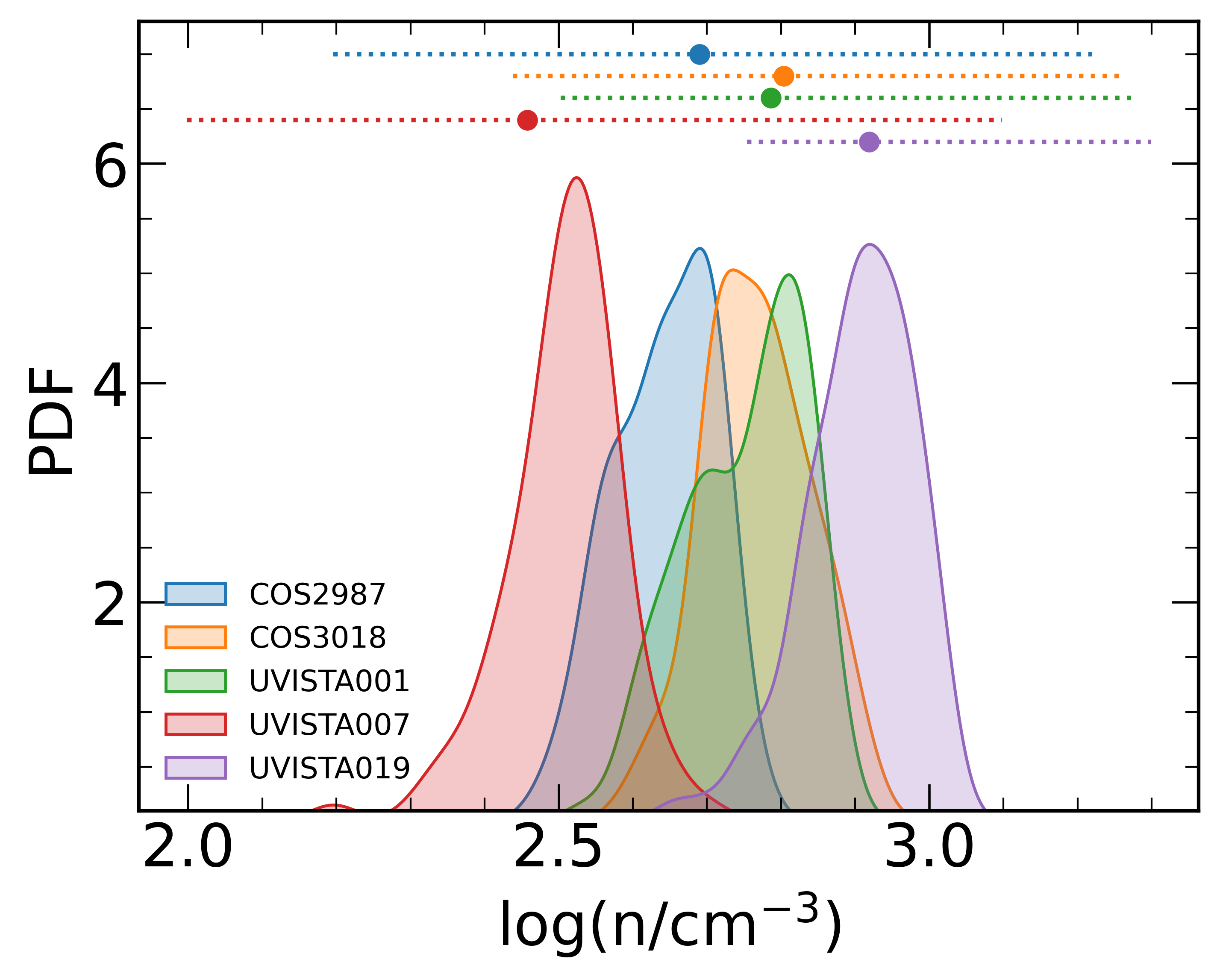}
    \includegraphics[scale=0.5]{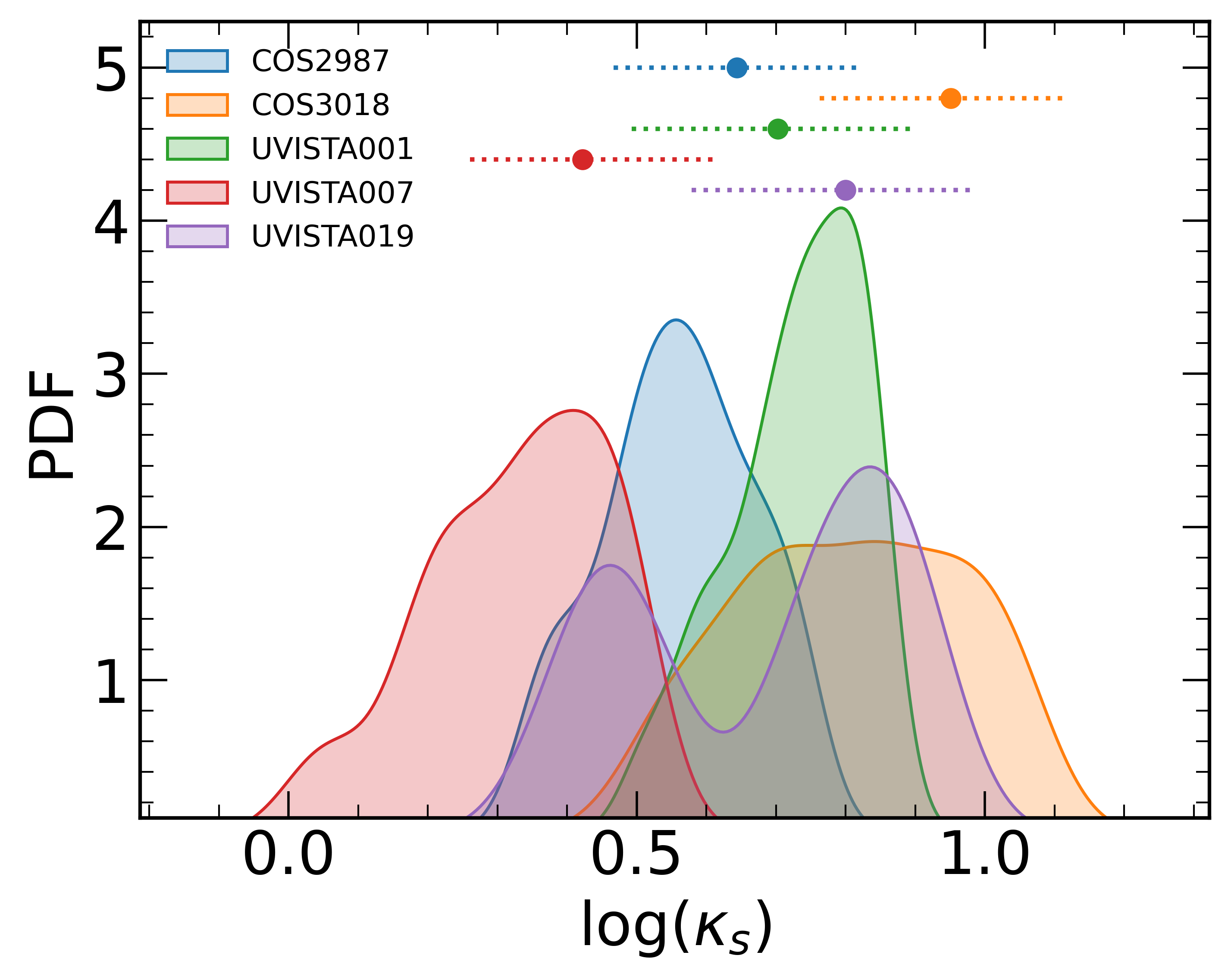}
    \includegraphics[scale=0.5]{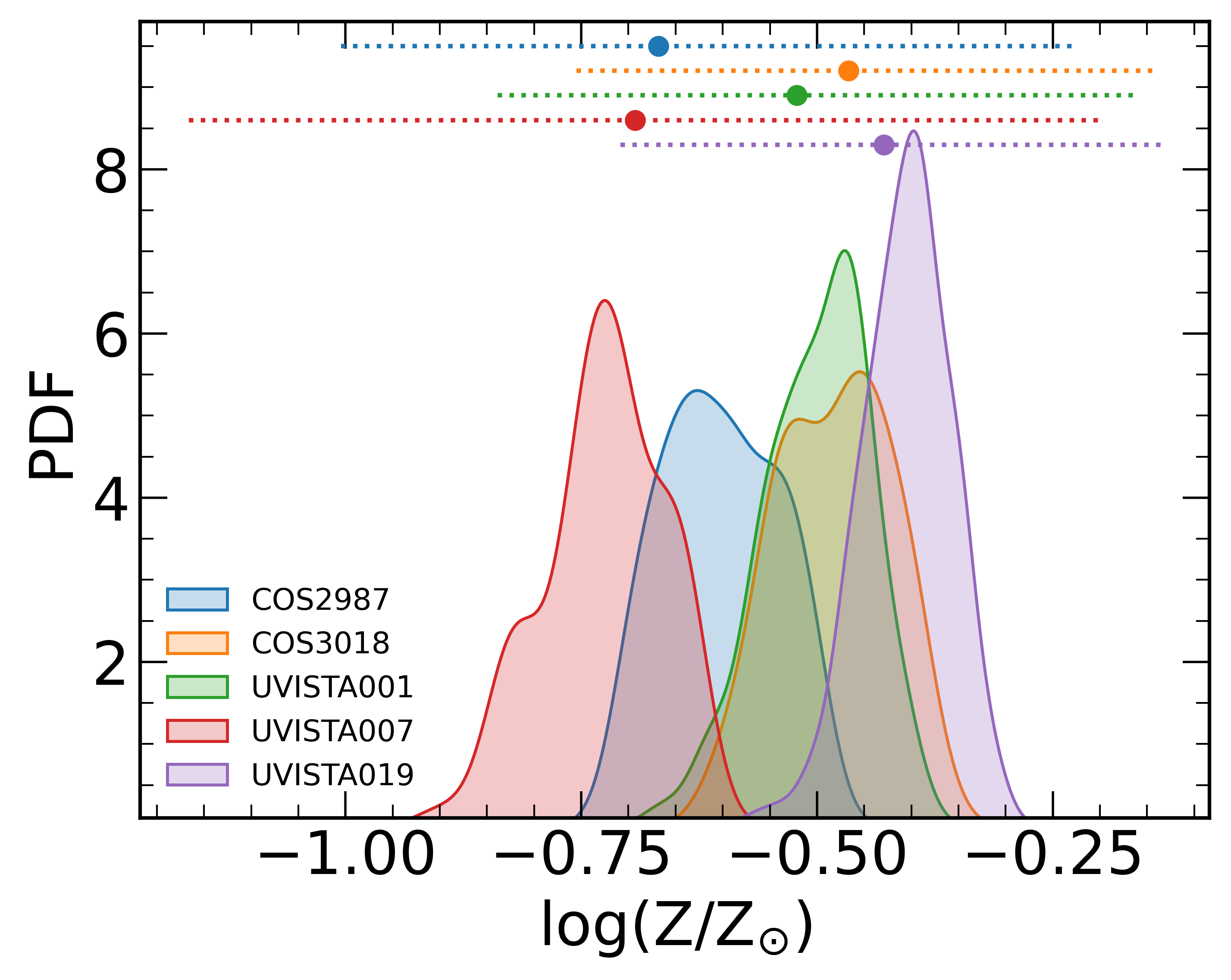}
    \caption{Probability density distribution (PDF) of the gas density ($n$, upper panel), deviation from the KS relation ($\kappa_s$, center), and metallicity ($Z$, lower panel) over the pixel maps in the five galaxies. The colored dots (dotted lines) represent the global value (uncertainty) for $n$, $\kappa_s$, $Z$ that derived with \texttt{GLAM} for each galaxy (same color code of the PDFs) when using the global $\Sigma_{\rm [CII]}$,  $\Sigma_{\rm [OIII]}$ and $\Sigma_{\rm SFR}$.}
    \label{fig:parameter_distributions}
\end{figure} 

In Figure \ref{fig:parameter_distributions} we analyze the probability distribution function (PDF) of $n$, $\kappa_s$, and $Z$ derived on pixel-by-pixel basis within the five galaxies. The relative error on the $(n,\,\kappa_s,\,Z)$ parameters, see Appendix \ref{appendix:error_maps} for the corresponding maps, are in the range $\Delta n/n \sim 0.3 - 0.5$, $\Delta \kappa_s/\kappa_s \sim 0.15-0.20$, $\Delta Z/Z \sim0.2 -0.5$, respectively, depending on the source. The density distribution in all the galaxies is fairly narrow ($\approx 0.4$ dex between the minimum and maximum value) and the peak of the distribution ranges from $\log(n/\rm cm^{-3})=2.5$ (UVISTA-Z-007), to $\log(n/\rm cm^{-3})=2.9$ (UVISTA-Z-019), albeit higher resolution data might reveal larger variability in the density within the ISM of the sources. For comparison, we also report the \textit{global value} and uncertainties for the same parameters obtained using the mean $\Sigma_{\rm [CII]}$, $\Sigma_{\rm [OIII]}$, and $\Sigma_{\rm SFR}$ of each source. The global $n$ (see Table \ref{tab:properties}) for each source is very close to the peak of the corresponding PDF over the pixels.
This implies that using \texttt{GLAM} for deriving the gas density of a galaxy using the average \CII, \OIII, and SFR surface density would return values that are representative of the actual ISM conditions within the source.\\

Our derived gas densities are slightly higher than the electron density, $\log (n_{e}/\rm cm^{-3})\approx 2.2$, inferred by \citet{fujimoto2022} using \OIII~$88\, \rm \mu m$ and \OIII~$\lambda$5007 JWST data in a $z\approx8.4$ source. This is expected because the electron densities derived with standard methods based on optical/UV line ratios 
\citep[e.g.][]{kewley2019} are sensitive to the conditions in the \HII~regions only, which despite being connected with the surrounding environment, have an overall lower density than the neutral/molecular gas in the PDRs \citep[Fig. 1 in][]{vallini2021}.
It is also interesting to compare our results with those obtained by \citet{davies2021}
on the redshift evolution of the (electron) density. \citet{davies2021} find an increasing trend with redshift from $\log (n_e/\rm cm^{-3})\approx 1.5 $ at $z\approx 0$ to $\log (n_{e}/\rm cm^{-3})\approx 2.4$ at $z\approx 2.3$. Such a positive correlation is likely connected with the evolution in the normalization of the star formation main sequence, and with the \HII~regions being embedded in parent giant molecular clouds (GMCs) characterized by higher densities at high-$z$ \citep[see also][]{sommovigo2020}. Note that the total gas density derived with \texttt{GLAM} is expli\-ci\-tly lin\-ked to that of the GMCs as, by construction, the density is parametrized in term of the 
\HII~region-PDR complexes tracing \OIII~and \CII, respectively. 

Our results at $z\approx7$ are also in agreement with the density increase ($n_e\geq 300 \rm \, cm^{-3}$) with redshift recently found by \citet{isobe2023} exploiting [OII]$\lambda \lambda$3726,3729 fluxes in $z\approx4.0-9$ sources. \citet{isobe2023} identify an increase of the electron density with redshift that can be approximated as $n_e\propto(1+z)^{p}$, with $p\sim1-2$. The exponent is explained by a combination of the compact morphology toward high-$z$, and the reduction of the electron density due to high electron temperatures of high-$z$ metal-poor nebulae. Our method favors the $p\approx 2$ solution that implies $n_e\approx500-1000\rm \, cm^{-3}$ at $z\approx 7$. \\

\begin{figure}
    \centering
    \includegraphics[scale=0.5]{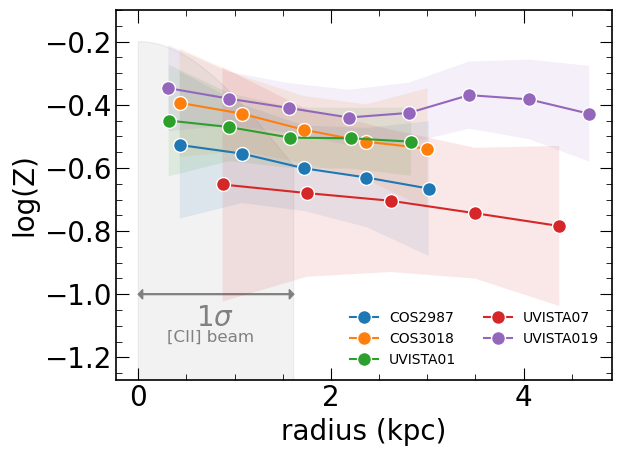}
    \caption{Radial profiles of the gas metallicity $Z$ for the five sources analyzed in this work. The gray shaded region denotes the 1$\sigma$ width of the median \CII~beam. The shaded colored regions represent the $1\sigma$ error, see the text for details on the calculation.}
    \label{fig:radial_profiles}
\end{figure} 

 All five galaxies lie above the KS relation, i.e. they are characterized by $\kappa_s>1$. Overall the $\kappa_s$ PDFs span a range between $\log(\kappa_s)\approx 0.3$ (UVISTA-Z-007), up to $\log(\kappa_s)\approx 1.3$ (UVISTA-Z-019). Note that for UVISTA-Z-019 we recover a clear bi-modality in the PDF of the burstiness parameter. This is because the dust-continuum detected region is more bursty than the outer part that is instead undetected with ALMA in continuum at $\approx 160\mu m$. As for the gas density, the $\kappa_s$ derived from global values is close to the peak of the corresponding PDF over the pixels, albeit the PDFs show a larger scatter, while the error on the global $\kappa_s$ is rather small ($\approx 0.2$ dex). Moreover, the global $\kappa_s$ tends to be skewed towards the higher end of the PDFs (see the case of UVISTA-Z-019).
\begin{figure*}
    \centering
    \includegraphics[scale=0.75]{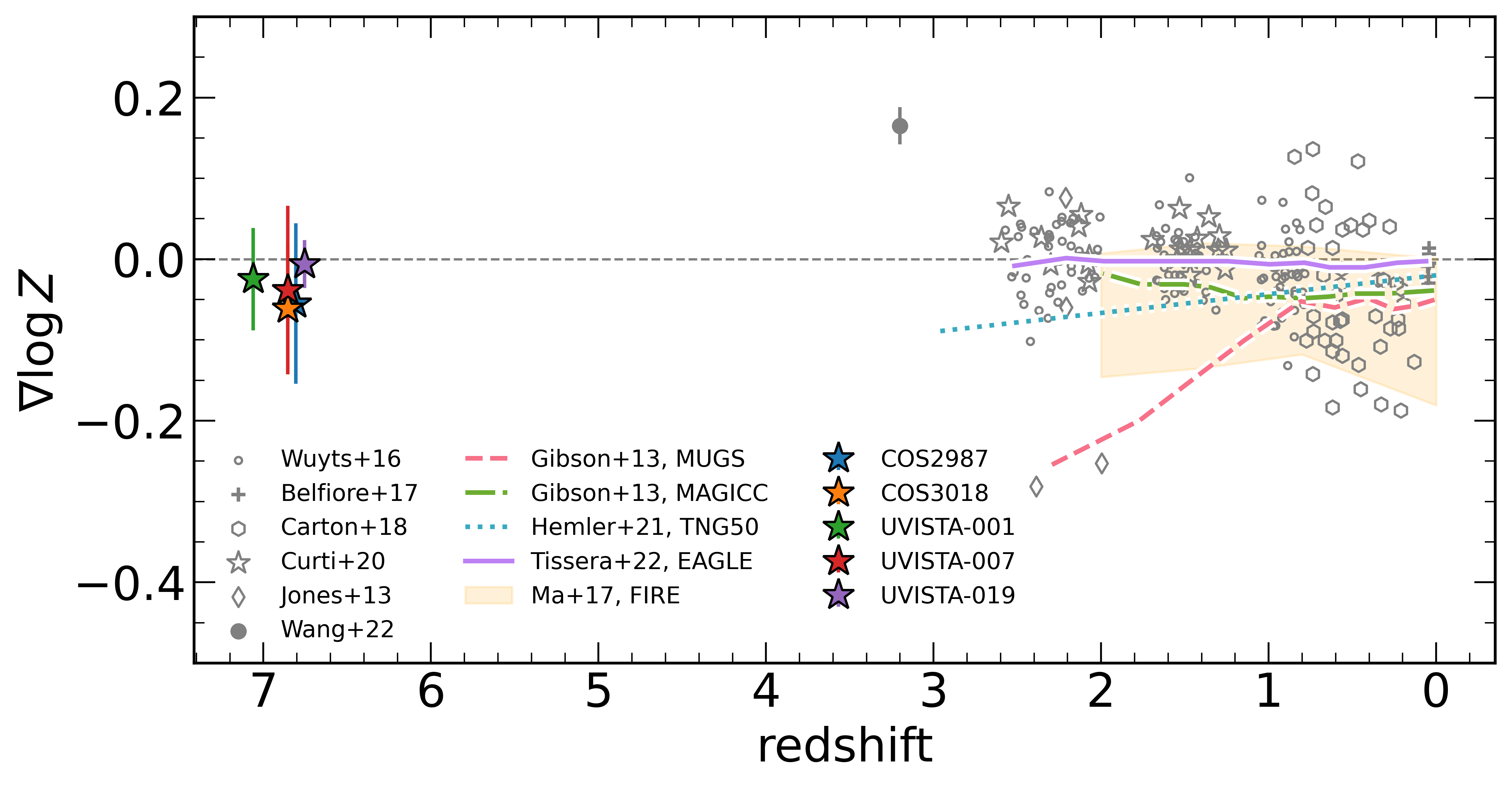}
    \caption{The redshift evolution of the metallicity gradient. Data, plotted without errors for {\bf clarity}, are taken from \citet[][stars]{curti2020}, \citet[][hexagons]{carton2018}, \citet[][crosses]{belfiore2017}, \citet[][small circles]{wuyts2016}, \citet[][diamonds]{jones2013}, \citet[][filled circle]{wang2022}. Simulated gradients taken from the
   literature are shown with lines, \citet[][pink dashed line: MUGS, normal feedback, green dot-dashed line: MAGICC enhanced feedback]{gibson2013}, \citet[][IllustrisTNG, cyan dotted line]{hemler2021}, \citet[][EAGLE, purple solid line]{tissera2022MNRAS}, \citet[][FIRE, orange shaded area]{ma2017}. The gradients inferred for our five sources are marginally negative but consistent with a slope equal to 0.}
    \label{fig:gradient_evolution}
\end{figure*}

Finally, we report the PDFs of the gas-phase metallicity obtained with \texttt{GLAM} in every pixel. As previously discussed for the density, and burstiness parameter, also the global $Z$ of each galaxy is close to the peak of the corresponding PDFs. The uncertainties in the global values are much larger than the typical width of the PDFs and comparable to the pixel relative error.All the galaxies have sub-solar gas metallicities ranging from $\log Z/Z_{\odot}=-0.75$ of UVISTA-Z-007 up to the $\log(Z/Z_{\odot})=-0.25$ in some of the pixels of UVISTA-Z-019. As a caveat we stress that the \texttt{GLAM} model assumes a fixed O/C ratio; the O/C in the early Universe is expected to be higher than that measured at $z=0$ \citep[e.g.][]{maiolino2019} thus the $Z$ inferred from the (high) \Soiii/\Scii~ratios might be underestimated by \texttt{GLAM}.

\subsection{Metallicity gradients}\label{subsec:radial_prof}
Metallicity gradients are sensitive probes of the complex network of processes that regulates gas inflows/outflows, feedback, and mixing within the ISM of galaxies across cosmic time \citep[e.g.][]{sanchezalmeida2014}. Theoretical models and numerical simulations \citep[e.g.][]{gibson2013, ma2017, hemler2021, sharda2021, tissera2022MNRAS} addressed the physical mechanisms shaping the metallicity gradient within galaxies and its evolution with redshift. In particular, negative gradients represent one of the strongest pieces of evidence for the inside-out galaxy formation scenario in which the nucleus forms first and more metals enrich the galaxy centre as compared to the disc. A flattening in the gas-phase galaxy metallicity can instead be the signature of star formation and strong stellar feedback \citep[e.g.][]{gibson2013, ma2017}, merger events \cite[e.g.][]{rupke2010} and pristine gas inflows towards the
central regions \citep[e.g][]{ceverino2016}. The combined effects of these physical
processes modulate the evolution of the metallicity gradients as function of redshift.

Thanks to the pixel-by-pixel derivation of the ISM properties within the five galaxies in our sample, it is possible to study how the gas metallicity, but also the galaxy burstiness and gas density, vary within each source as a function of the galactocentric radius. 
To do so we use the \texttt{RadialProfile} class within the \texttt{photutils} package to compute the azimuthally-averaged value, and the corresponding uncertainty, of all the three \texttt{GLAM} parameters over circular annuli of $\approx0.8$ kpc (2 pixels) width. The exception is UVISTA-Z-007 for which the radial spacing corresponds to 1 pixel, due to the lower spatial resolution of the data. We assume the center ($r=0$) to be located at peak of the \OIII~emission. The error on the radial profiles is computed by providing to the \texttt{RadialProfile} routine the map of the $1 \sigma_i$ errors of each parameter in each $i$-th pixel as computed by \texttt{GLAM} (see Appendix \ref{appendix:error_maps}).
\begin{table*}
\centering
\begin{tabular}{lccccccc}
\hline
\hline
      name &  redshift &           $\log(n/{\rm {cm^{-3}}})$ &           $\log(\kappa_s)$&             $\log (Z/Z_{\odot})$& $\nabla (\log Z)$ & $t_{\rm dep}\, \rm (Myr)$\\
\hline
   COS-2987 &     6.807 & 2.69$_{-0.49}^{+0.53}$ & 0.64$_{-0.18}^{+0.18}$ & -0.67$_{-0.34}^{+0.44}$ & $-0.05 \pm 0.09$ & $190 \pm 56$\\
   COS-3018 &     6.854 & 2.80$_{-0.37}^{+0.46}$ & 0.95$_{-0.19}^{+0.17}$ & -0.47$_{-0.29}^{+0.33}$ & $-0.06 \pm 0.08$ & $100 \pm 40$\\
UVISTA-Z-001 &     7.060 & 2.79$_{-0.28}^{+0.49}$ & 0.70$_{-0.21}^{+0.19}$ & -0.52$_{-0.32}^{+0.36}$ & $-0.03 \pm 0.06$ & $96 \pm 21$\\
UVISTA-Z-007 &     6.749 & 2.46$_{-0.46}^{+0.64}$ & 0.42$_{-0.16}^{+0.20}$ & -0.69$_{-0.47}^{+0.49}$ & $-0.04 \pm 0.10$ & $280 \pm 130$\\
UVISTA-Z-019 &     6.754 & 2.92$_{-0.16}^{+0.38}$ & 0.80$_{-0.22}^{+0.19}$ & -0.43$_{-0.28}^{+0.29}$ & $~0.00 \pm 0.03$ & $84 \pm 14$\\
\hline
\end{tabular}
\caption{Derived physical properties for the galaxies analyzed in this work. We list the galaxy names (column 1), redshift (column 2), global values (see text for details) for the gas density (column 3), burstiness (column 4), and gas metallicity (column 5). In column 6 we report the radial gradient $\nabla (\log Z)$. Finally, in column 7 we report the median value of the depletion time over the spatially resolved pixels, and its $\pm \sigma$ deviation.}
\label{tab:properties}
\end{table*}
The result of this procedure is outlined in Figure \ref{fig:radial_profiles} where we show the gas metallicity profiles for the five sources analyzed in this work. We refer the interested reader to Appendix \ref{appendix:app_a} for details regarding the radial profiles of the gas density and burstiness. We note that $Z$ gradients as a function of galactocentric radius are marginally negative, but consistent with being flat within the errors. The median in the sample $\nabla (\log Z)\approx -0.03 \pm 0.07$ dex/kpc and the values for each source can be found in Table \ref{tab:properties}.
Our analysis suggests that the metal enrichment in the central regions might be connected with recent burst (high $\kappa_s$) of star formation and that, at the same time, the vigorous starburst \citep[and possibly flikering SFR, e.g.][]{pallottini2023} implies copious energy injection via stellar feedback, thus flattening the gradient.

In Figure \ref{fig:gradient_evolution}, we report the gradient estimates for the five sources as a function of their redshift, comparing them with observations and theoretical models in the literature. Negative metallicity gradients in the radial direction have been confirmed in most galaxies at $z\approx 0$ \citep[e.g.][]{stanghellini2010, belfiore2017, stanghellini2018} whereas there is evidence of an evolution of the metallicity gradients towards a flattening at high-$z$ \citep{wuyts2016, carton2018, curti2020} albeit with large scatter \citep[see e.g. the negative slopes reported by][]{jones2013}. 
 
In the coming years, high-$z$ samples with spatially resolved metallicity measurements will rapidly expand thanks to JWST that will allow observing nebular lines (e.g. [O III]$\lambda5007$, H$\beta$, [O II]$\lambda \lambda 3726, 3729$) that are routinely used as metallicity estimators. \citet{wang2022} reported the first JWST determination of the metallicity gradient in a $z\approx 3.2$ galaxy, finding a strongly positive gradient likely due to the interaction with a nearby object. With our method, we do not find evidence of positive gradients in our sources at $z\approx 7$ and this is likely connected to the fact that the star formation burst, as traced by \OIII, is localized at the center of the sources. Nevertheless, as a caveat, we warn the reader that our derivation of the metallicity using \texttt{GLAM} represents an indirect methodology and thus the uncertainty in the comparison with gradients in the O/H abundance obtained with other methods is large. {\color{black} In particular, systematics in the $Z$ determination can arise from the fact that the likely enhancement O/C ratio at low metallicity \citep[e.g.][]{maiolino2019} is not accounted for in GLAM (see Section \ref{sec:model}). This assumption implies that the $Z$ inferred with GLAM from the $\Sigma_{\rm [OIII}$ and $\Sigma_{\rm [CII]}$ fitting, might be biased towards higher values to compensate for a lower than expected oxygen abundance at sub-solar metallicity.}

\subsection{Kennicutt Schmidt relation and gas depletion time}
 \label{subsec:kennicutt_schmidt}
\begin{figure*}
    \centering
    \includegraphics[scale=0.7]{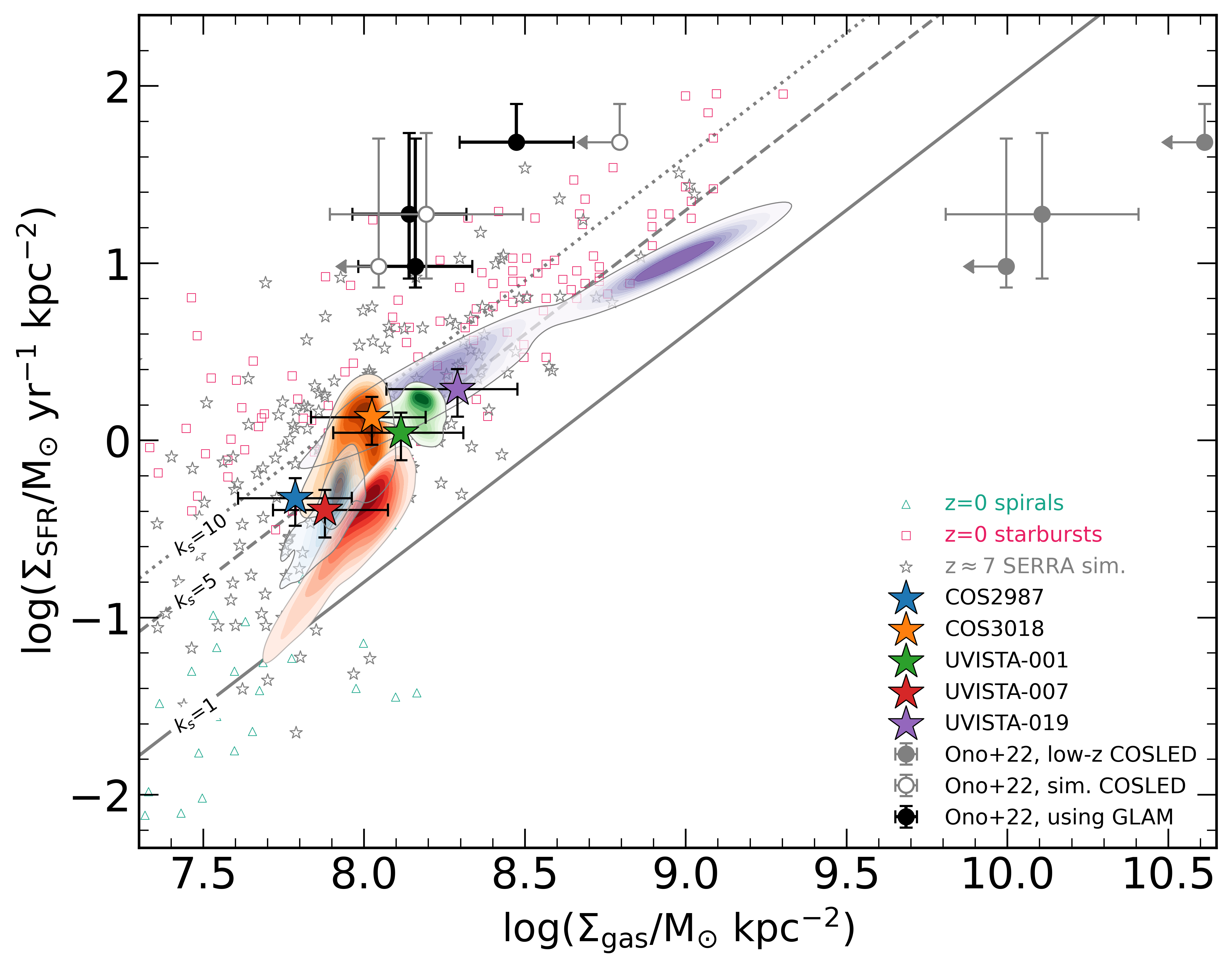}
    \caption{Relation between SFR and total (atomic plus molecular) gas surface densities $\Sigma_{\rm SFR} = \kappa_s 10^{-12} \Sigma^{1.4}_{\rm gas}$ \citep{heiderman10}. The solid (dashed, dotted) line represents the relation for $\kappa_s=1$ ($\kappa_s=5$, $\kappa_s=10$, respectively). Local spiral galaxies \citep{delosreyes2019} and starburst galaxies \citep{kennicutt2021} are indicated with green triangles and magenta squares, respectively. 
    For the local spiral (starburst) galaxies the molecular component of the total gas surface density is derived assuming the Milky Way CO-to-H$_2$ conversion factor $\alpha_{\rm CO, MW}=4.3\, \aco$  (ULIRG conversion factor $\alpha_{\rm CO, ULIRG}=0.86 \, \aco$). 
    Simulated $z\approx7$ galaxies from \code{SERRA} \citep{pallottini2022} are indicated with empty gray stars. The 2D density distribution of $\Sigma_{\rm SFR}$ vs $\Sigma_{\rm gas}$ of the pixels within each of the five $z\approx7$ galaxies analyzed in this work are represented with shaded colored regions. The colored stars indicate the location of the sources when considering their global values. We complement the plot with three LBGs at $z\approx 6$ that have been detected in \CII, \OIII~\citep{harikane2020}, and recently followed up in CO(6--5) by \citet{ono2022}. Their location (grey filled vs grey empty circles) depends on the assumed CO excitation (low-$z$ COSLED vs high-$z$ simulated COSLED) and CO-to-H$_2$ conversion factor ($\alpha_{\rm CO, MW}$ vs $\alpha_{\rm CO}=1.4\, \aco$ derived by \citet{vallini2018} in the pilot \code{SERRA} simulation \citep{pallottini2017b}). The location in the KS plane of the three LBGs derived from \CII~and \OIII~data using \texttt{GLAM} (\citetalias{vallini2021}) is indicated with black circles.}
    \label{fig:KS_relation}
\end{figure*}

In Figure \ref{fig:KS_relation} we show the $\Sigma_{\rm SFR}$-$\Sigma_{\rm gas}$ (Kennicutt-Schmidt) relation for the five galaxies analyzed in this work. We consider both the 2-D probability density distribution of the values within the pixels, and the location of the five galaxies in the KS plane when deriving the $\Sigma_{\rm gas}$ from the global $\kappa_s$ and $\Sigma_{\rm SFR}$. 

In line with previous works (\citetalias{vallini2021}), we confirm that the inferred location in the KS plane of a source, obtained considering the global \Scii~and \Soiii~values, traces the most starbursting (high $\kappa_s$) patches within the ISM of the objects. The global values fall indeed in the upper-$\kappa_s$ end of the 2-D distribution of the pixels within each galaxy.
 The global and spatially resolved \Sg-\S* values place our galaxies above the $z=0$ relation for spiral galaxies \citep{delosreyes2019} in the region populated by starburst sources in the local Universe \citep{kennicutt2021}. 
 
 Interestingly, their location is in good agreement with the position in the KS plane of simulated star forming galaxies at $z\approx7$ extracted from the \code{SERRA} zoom-in cosmological simulation \citep{pallottini2022} 
that cover a range of stellar masses ($10^8 \, {\rm M}_{\odot }\lesssim M_\star \lesssim 5\times 10^{10}\, {\rm M}_{\odot }$) and star formation rate (SFR$\approx 1-100\, \rm M_{\odot}\, yr^{-1}$) that encompass that of the five LBGs analyzed in this work. 

Our analysis suggests that luminous LBGs in the EoR are characterized by an efficient conversion of the gas into stars. To put this conclusion into a broader context we perform a comparison with three luminous $z\approx6$ LBGs (J0235-0532, J1211-0118 and J0217-0208) first targeted by \citet{harikane2020} with ALMA in \CII~and \OIII~characterized by UV luminosities ($\approx 3 \times 10^{11}\rm \, L_{\odot}$) and \OIII/\CII~ratios ($\approx 3-8$) similar to those of our galaxy sample \citep[see][]{witstok2022}.
\citetalias{vallini2021} studied J0235-0532, J1211-0118 and J0217-0208 with \texttt{GLAM}, albeit using only the barely resolved \CII~and \OIII~data available at the time, deriving their burstiness parameter, gas-phase metallicity and density. Thanks to the detection of the CO(6--5) line in J0235-0532, and the upper limits in the other two LBGs \citep{ono2022}, it is possible to compare their location in the KS plane from \texttt{GLAM} with that obtained using the CO as fiducial gas proxy, which, however, involves many uncertain parameters. In fact, (i) the conversion of the CO(6--5) flux into the CO(1--0)\footnote{The CO(1--0) luminosity can be then converted into the molecular mass via the CO-to-H$_2$ conversion factor, $M_{mol}=\alpha_{\rm CO} L'_{\rm CO(1-0)}$} depends on the CO Spectral Line Energy Distribution (COSLED) excitation, which is observationally unconstrained in LBGs in the EoR \citep[][]{pavesi2019}, (ii) the CO-to-H$_2$ conversion factor is also highly uncertain in high-$z$ sources as it depends on metallicity \citep{bolatto2013} and (iii) the actual size of the molecular gas distribution cannot be derived from unresolved CO observations. As discussed by \citet{ono2022}, assuming $\alpha_{\rm CO, MW}=4.3\, \aco$ and the CO(6--5)/CO(1--0) ratio of $z\approx 1$ star forming galaxies \citep{daddi2010}, all the three sources lie below the KS relation, at odds with their high \OIII/\CII~ratios that are usually powered by ongoing bursts of star formation \citep{arata2020, vallini2021, kohandel2023}. If, instead, one assumes the $\alpha_{\rm CO}=1.4 \, \aco$ and CO(6--5)/CO(1--0) ratio derived by \citet{vallini2018} for Althaea --  a simulated LBG extracted from the precursor of the \code{SERRA} zoom-in simulation \citep{pallottini2017b, pallottini:2019, pallottini2022} -- the sources are compatible with the location of starburst galaxies in agreement within the errors (see Fig. \ref{fig:KS_relation}) with the location inferred with \texttt{GLAM} in \citetalias{vallini2021}.
Given that the CO(6--5) traces dense/warm ($n_{crit}\approx 3 \times 10^5\rm cm^{-3}$, $T_{\rm ex}\approx 116$ K) molecular gas \citep[][for a recent review]{wolfire2022} the line is expected to be luminous in bursty galaxies that are experiencing on-going star formation within dense GMCs \citep[][]{vallini2018}. UVISTA-Z-019 would be therefore an ideal target for CO follow-up being the most dense and bursty among the five LBGs in our sample with both CO(6--5) and CO(7--6) falling into ALMA band 3. Moreover, its continuum detection translates into SFR$_{\rm IR}$/SFR$_{\rm TOT}=0.7$, namely 70 per cent of the star formation is dust obscured and dust shielding is one of the key necessary conditions mitigating the CO dissociation in star forming regions \citep{wolfire2010}. Note that the UVISTA-Z-019 IR luminosity (and the obscured fraction of the star formation) $L_{\rm IR}=3.1 \times 10^{11}  L_{\odot}$, (SFR$_{\rm IR}$/SFR$_{\rm TOT}=0.7$) is similar to that of J0235–0532 ($L_{\rm IR}=5.8 \times 10^{11} L_{\odot}$, $\rm SFR_{IR}/SFR_{TOT}=0.6$, respectively), the CO(6-5)-detected LBG from the \citet{ono2022} sample.\\

\begin{figure}
    \centering
    \includegraphics[scale=0.4]{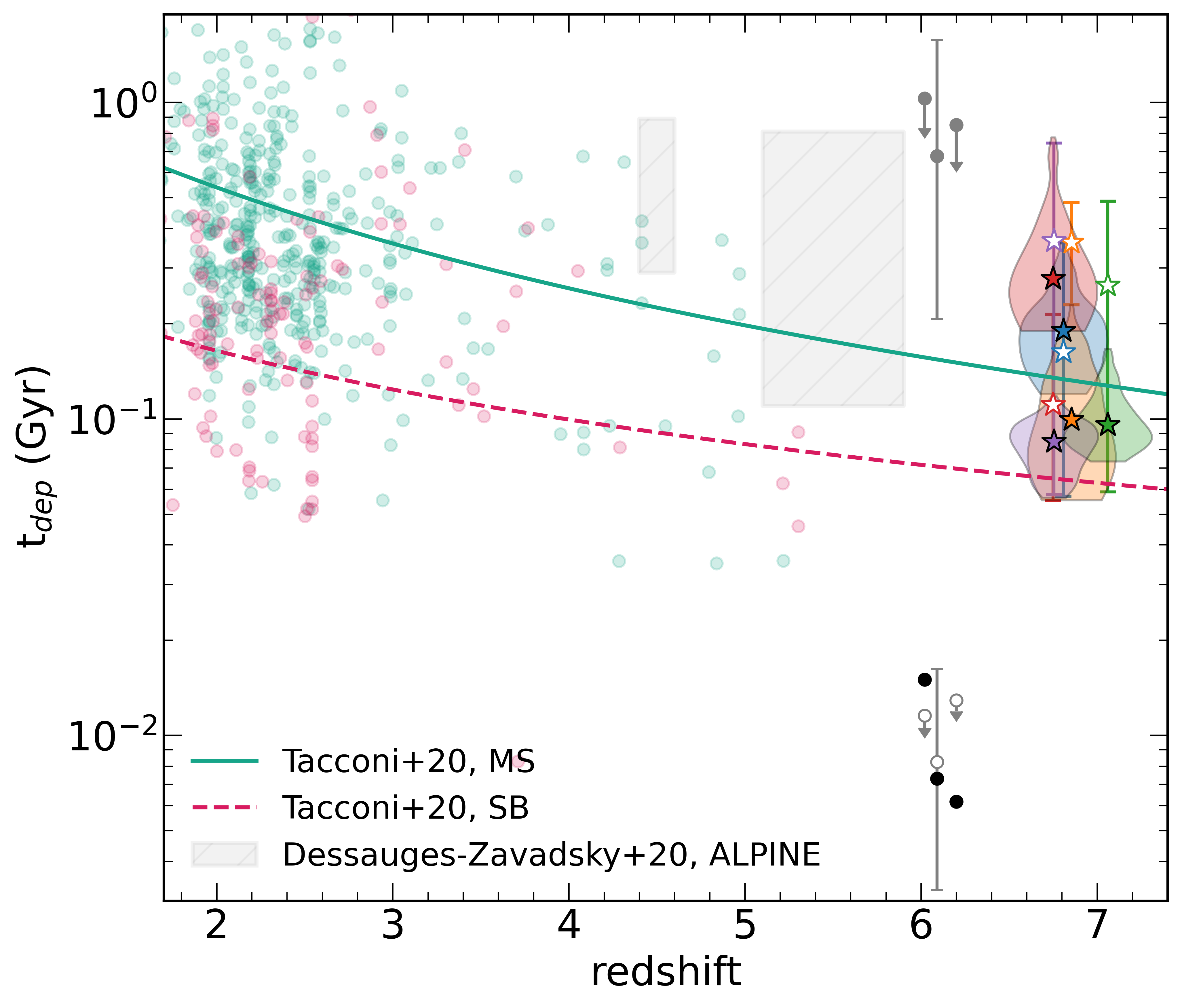}
    \caption{Redshift evolution of the depletion time. The colored violin plots represent the distribution of $t_{\rm dep}$ within our galaxies as inferred from the spatially resolved $\kappa_s$. The median value (also reported in Table \ref{tab:properties}) is highlighted with a filled colored star. {\color{black} For comparison, empty stars with color bars represent the integrated depletion time ($t_{\rm dep}=M_{\rm gas}/\rm SFR$) derived using the \citet{zanella2018} relation for inferring the gas mass from the \CII~ luminosiy, and considering the SFR$_{\rm tot}$ by \citet{witstok2022}. We adopt the same color-code of the previous figures.} The extrapolation out to $z=7.5$ of the best fit relations from \citet{tacconi2020} for main sequence galaxies (green points) and starburst galaxies (magenta points) are indicated with solid and dashed lines, respectively. The $t_{\rm dep}$ from ALPINE in two redshift bins \citep{dessauges-zavadsky2020} is indicated with gray shaded areas. The $t_{\rm dep}$ inferred by \citet{ono2022} for the \citet{harikane2020} LBGs is indicated with filled and empty gray circles (same color-code of Fig. \ref{fig:KS_relation}) while that inferred using \texttt{GLAM} with filled black circles.}
    \label{fig:tdep_evolution}
\end{figure}

High burstiness parameters translate into short gas depletion times \citep[e.g.][]{tacconi2013, tacconi2020}, defined on spatially resolved scales as $t_{\rm dep}=\Sigma_{\rm gas}/\Sigma_{\rm SFR}\propto \kappa_s^{-1} \Sigma_{\rm gas}^{-0.4}$. The evolution of the depletion time with redshift (see Figure \ref{fig:tdep_evolution}) is a fundamental quantity shaping galaxy evolution, as it quantifies the typical timescales for the conversion of the gas into stars, thus ultimately the galaxy and stellar build up from the Dark Ages to the present day and the efficiency with which galaxies build up their stellar mass. In particular, \citet{tacconi2020} found that the integrated depletion timescale, {\color{black}namely that derived as $t_{\rm dep}=M_{\rm gas}/\rm SFR$}, depends mainly on the redshift and offset
from the main sequence ($t_{\rm dep}\propto (1 + z)^{-1} \times \Delta_{MS}^{-0.5}$). The trend for main sequence galaxies has been overall confirmed in the $z\approx 4.5 - 5.8$ redshift range by the ALPINE \citep{lefevre2020} survey \citep{dessauges-zavadsky2020} {\color{black}by inferring the gas mass from the \CII~luminosity \citep[see][]{zanella2018}}. 

From the spatially resolved $\kappa_s$ within our sources we obtain median $t_{\rm dep}$ ranging from $\approx280$ Myr of UVISTA-Z-007 to $\approx 80$ Myr of UVISTA-Z-019, in agreement with the redshift evolution proposed by \cite{tacconi2020}. In particular, these values encompass the tight range between the extrapolation of $t_{\rm dep}$ out to $z\approx 8$ for main sequence galaxies and that for starburst (deviation from the main sequence, $\Delta_{MS}=10$) sources.

{\color{black} Finally, we also compare the spatially resolved $t_{\rm dep}$ obtained from \texttt{GLAM} against that derived on global scales by using the \citet{zanella2018} conversion factor to infer the gas mass $M_{\rm gas}=\alpha_{\rm [CII]} L_{\rm [CII]}$. This also enables a fair comparison with the depletion times in ALPINE \citep{dessauges-zavadsky2020}, which appear still marginally higher than those derived in our five galaxies.\\ 
For COS-2987 the global value of the $t_{\rm dep}$ is in excellent agreement the median of the spatially resolved one. For the other four sources the two methods return $t_{\rm dep}$ that agree within the errors (see Fig. \ref{fig:tdep_evolution}). More precisely, in COS-3018, UVISTA-Z-001, UVISTA-Z-019 the global value is higher than the median of the spatially resolved one. Overall we interpret this trend as the probe that \texttt{GLAM}, by using both the information on \CII~and \OIII, is more sensitive to the starbursting regions within galaxies, whose depletion time is expected to be shorter than the average value. The exception is UVISTA-Z-007 that is the galaxy with the poorer spatial resolution of the \CII~and \OIII~data for which the median of the spatially resolved value is lower than the global $t_{\rm dep}$}.
We point out that thanks to the derivation of the $t_{\rm dep}$ with \texttt{GLAM} the five galaxies analyzed here allow to constrain the extrapolation of the redshift evolution of the depletion time in the EoR, and the method is definitively promising as an alternative in galaxies for which the CO detection might be challenging.

\subsection{Linking dust temperature and gas depletion time}
\label{subsec:dust}
\begin{figure}
    \centering
    \includegraphics[scale=0.36]{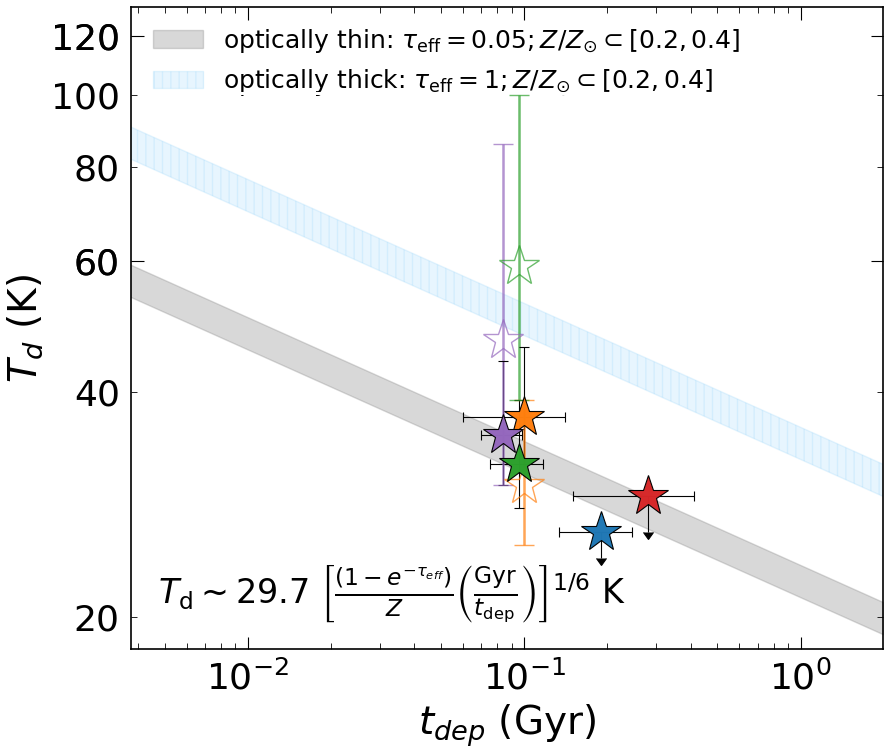}
    \caption{The dust temperature ($T_d$) as a function of the depletion time ($t_{\rm dep}$), the color-code is the same as in previous Figures. Filled stars represent $T_d$ derived with the \citet{sommovigo2021} method, while empty stars those obtained by \citet{witstok2022} by fitting the SED (probed by 2 photometric points) for the five LBGs analyzed in this paper. The theoretical relation between $T_{\rm d}$ and $t_{\rm dep}$ is shown in the lower left corner. The gray (blue) shaded region highlights the solutions for $\tau_{\rm eff}=0.05$ ($\tau_{\rm eff}=1$) and varying the metallicity in the range spanned by our estimates with \texttt{GLAM} for our sources.}
    \label{fig:td_ks}
\end{figure}

In the last few years, several works dealing with FIR stacked SED fitting across cosmic time ($z\approx 0-10$), have inferred the presence of a tight correlation between $T_{\rm d}$ and redshift (\citealt{Schreiber18,Bouwens20,Viero22}, see also \citealt{Liang19}). \cite{sommovigo2022a,sommovigo2022b} proposed a theoretical explanation for the $T_{\rm d}-z$ relation based on the evolution of the total gas depletion time. They show that, in a simplified single-phase ISM model, $T_{\rm d} \propto t_{\rm dep}^{-1/6}$ (see eq. 10 in \citealt{sommovigo2022a} also reported in Fig. \ref{fig:td_ks}), resulting in  a mild increase of $T_{\rm d}$ with redshift ($T_{\rm d}\propto (1+z)^{0.4}$) due to the shorter $t_{\rm dep}$ at early epochs (Fig. \ref{fig:tdep_evolution}) produced by the more vigorous cosmic accretion. At any fixed epoch, the scatter in $T_{\rm d}$ is produced by variations in metallicity and optical depth $\tau_{\rm eff}$ (see Figure \ref{fig:td_ks}), with lower $Z$ (higher $\tau_{\rm eff}$) resulting in warmer dust. 

To investigate this scenario, in Figure \ref{fig:td_ks} we report the dust temperature for the sources in our sample as a function of the depletion times derived in the previous Section. Unfortunately, for most of our sources (and in general for galaxies towards the EoR, e.g. \citealt{bethermin2020, gruppioni2020,inami2022}), only a single/two ALMA dust continuum detections are generally available. Thus, the $T_{\rm d}$ derived from SED fitting is highly uncertain ($\Delta T_{\rm d}/T_{\rm d} \simgt 60\%$, see \citealt{witstok2022}) hampering an unambiguous analysis. 

To overcome this problem, and study in detail the $T_d-t_{\rm dep}$ relation in all the sample, we use an alternative method \citep{sommovigo2021} based on the combination of the ALMA FIR continuum data point with the \CII~luminosity information. The latter is used as a proxy of the total gas (and dust) mass, so that the single continuum measurement can be exploited to constrain the dust temperature. In this case this allows us to constrain $T_{\rm d}$ to a greater precision ($\Delta T_{\rm d}/T_{\rm d} \simgt 20\%$). We obtain $T_{\rm d}\approx 35\ \mathrm{K}$ for the $3$ dust-continuum detected galaxies, and $T_{\rm d}\lesssim 30\ \mathrm{K}$ for the $2$ continuum-undetected ones. These values are consistent within the uncertainties with those derived from the SED fitting by \citet{witstok2022}, where available.

As shown in Fig. \ref{fig:td_ks} we find that galaxies with shorter depletion times host warmer dust. Our results are in agreement with the physically motivated prediction for the optically thin ($\tau_{\rm eff} =0.05$) case \citep{sommovigo2022a,sommovigo2022b} when accounting for the metallicity variation within the sample. 

If future ALMA dust continuum observations at shorter wavelengths will confirm the warmer $T_{\rm d}$ suggested by the median values for UVISTA-Z-001 and UVISTA-Z-019 by \citet{witstok2022}, this might indicate that these galaxies are characterized by $\tau_{\rm eff}=1$ and thus a spatially segregated scenario between dust and UV emission. This can be further confirmed by high spatial resolution ALMA observations (tracing the dust obscured star formation) in conjunction with JWST (tracing the un-obscured one).

\section{Summary}
In this paper, we have used \texttt{GLAM} \citep{vallini2020,vallini2021} to derive the ISM properties (gas density, deviation from the KS relation, and gas metallicity) in five UV luminous LBGs at $z\approx7$ for which moderately resolved \CII~and \OIII~observations are available. We have compared the pixel-by-pixel values for the ISM parameters with the global ones derived with the same methodology using instead average \Scii~and \Soiii~surface brightness values. We confirm the conclusion by \citetalias{vallini2021}, namely that global values are biased towards the most luminous ISM regions. The main results from our spatially resolved analysis are the following:
\begin{itemize}
    \item The distribution of the gas density in the five LBGs is narrow and peaks in the range $\log (n/cm^{-3}) = 2.5 -3.0$, depending on the source. The gas densities obtained are higher than typical values in local galaxies, hence suggesting an overall increase in the mean gas density in the ISM at early epochs. 
    \item We derived radial profiles for the metallicity, density and burstiness. In particular, the metallicity shows a mildly negative radial gradient that, within the uncertainties, is compatible with being flat.
    \item All five galaxies lie above the KS relation by a factor of $\approx 3-10$, in perfect agreement with expectations from cosmological zoom-in simulations \citep{pallottini2022} at the same redshift. The $\kappa_s$ value is higher in the center. In some cases, we obtain a bimodal distribution in regions where dust continuum emission is detected, suggesting the presence of dense, dust-obscured, highly star-forming regions.
    \item We predict that bursty galaxies with dense gas (such as UVISTA-Z-019) would be an ideal target for ALMA follow-ups in CO(6--5) as mid-$J$ CO lines trace warm/dense molecular gas.
    \item The gas depletion times, derived from the KS relation, are in the range $t_{\rm dep}\approx 80-250$ Myr. The $t_{\rm dep}$ of the five sources fall between that predicted by the extrapolation out to $z\approx7$ of for MS and SB galaxies of the \citet{tacconi2020} relation.
    \item The dust temperature of the five sources correlates with  $t_{\rm dep}$., as predicted by theoretical models  \citep{sommovigo2021} for an optically thin medium. We confirm that the redshift evolution of the dust temperature might be the imprint of a more efficient conversion of the gas into stars.
\end{itemize}

The work presented in this paper highlights the huge potential of the synergy between physically motivated line emission models and spatially resolved observations of \CII~and \OIII~for constraining a wealth of ISM properties within galaxies in the Epoch of Reionization. Simply using a handful of observables -- traced down to kpc scales -- allows to determine the Kennicutt-Schmidt relation, metallicity profiles, and impact of gas accretion on the dust continuum properties. Further follow-ups at higher spatial resolution in a larger samples of sources already detected in \CII~will allow putting the results presented here on a statistically more robust basis. 

\label{sec:conclusions}

\section*{Data Availability}
\texttt{GLAM} can be accessed on Github at \url{https://lvallini.github.io/MCMC galaxylineanalyzer/}. HST data underlying this article are available in the MAST
archive at 10.17909/6gya-3b10 (GO 13793), 10.17909/T9-JHSFM392 (GO 16506) and from \url{https://archive.stsci.edu/prepds/3d-hst/} (the 3D-HST Treasury Program). The ALMA data are available in the ALMA science archive at \url{https://almascience.eso.org/asax/} under the following project codes: 2015.1.01111.S, 2018.1.01359.S, 2018.1.00429.S, 2018.1.01551.S, 2017.1.00604.S, 2015.1.00540.S, 2018.1.00085.S, 2018.1.00933.S, 2019.1.01611.S, 2019.1.01524.S,
2018.1.00085.S, 2019.1.01611.S. 

\section*{Acknowledgments}
LS, AF, MK acknowledge support from the ERC Advanced Grant INTERSTELLAR H2020/740120.
JW acknowledges support by Fondation MERAC, the Science and Technology Facilities Council (STFC), by the ERC through Advanced Grant 695671, ``QUENCH'', and by the UK Research and Innovation (UKRI) Frontier Research grant RISEandFALL.
AP acknowledges the CINECA award under the ISCRA initiative, for the availability of high performance computing resources and support from the Class B project SERRA HP10BPUZ8F (PI: Pallottini). 
SC acknowledges support by the European Union (ERC Starting Grant, WINGS, 101040227). Views and opinions expressed are however those of the author(s) only and do not necessarily reflect those of the European Union or the European Research Council Executive Agency. Neither the European Union nor the granting authority can be held responsible for them. RS acknowledges an STFC Ernest Rutherford Fellowship (ST/S004831/1).

This work was based on observations taken by the Atacama Large Millimeter/submillimeter Array (ALMA). ALMA is a partnership of ESO (representing its member states), NSF (USA) and NINS (Japan), together with NRC (Canada), MOST and ASIAA (Taiwan), and KASI (Republic of Korea), in cooperation with the Republic of Chile. The Joint ALMA Observatory is operated by ESO, AUI/NRAO and NAOJ. This work was furthermore partially based on new observations made with the NASA/ESA \textit{Hubble Space Telescope} (\textit{HST}), obtained at the Space Telescope Science Institute (STScI), which is operated by the Association of Universities for Research in Astronomy, Inc., under NASA contract NAS 5-26555. \textit{HST} archival data was obtained from the data archive at the STScI. STScI is operated by the Association of Universities for Research in Astronomy, Inc. under NASA contract NAS 5-26555.
We gratefully acknowledge computational resources of the Center for High Performance Computing (CHPC) at SNS.
This research made use of astropy \citep{astropy2018}, Matplotlib \citep{matplotlib2007}, Seaborn \citep{seaborn2020}, Scipy \citep{scipy2020}, Numpy \citep{NumPy2020}, and Photutils \citep{bradley2022}, an Astropy package for detection and photometry of astronomical sources.



\bibliographystyle{mnras}
\bibliography{bibliography}

\appendix
\section{Error maps}
\label{appendix:error_maps}
\begin{figure*}
    \centering
    \includegraphics[scale=0.4]{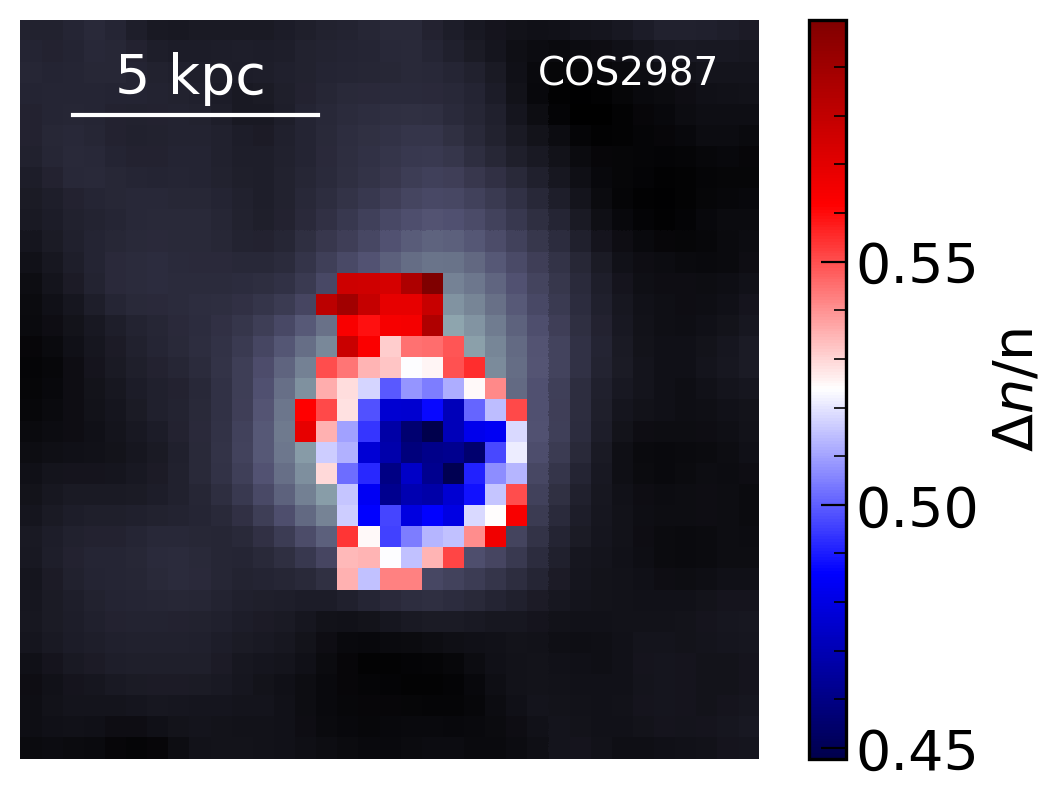}
    \includegraphics[scale=0.4]{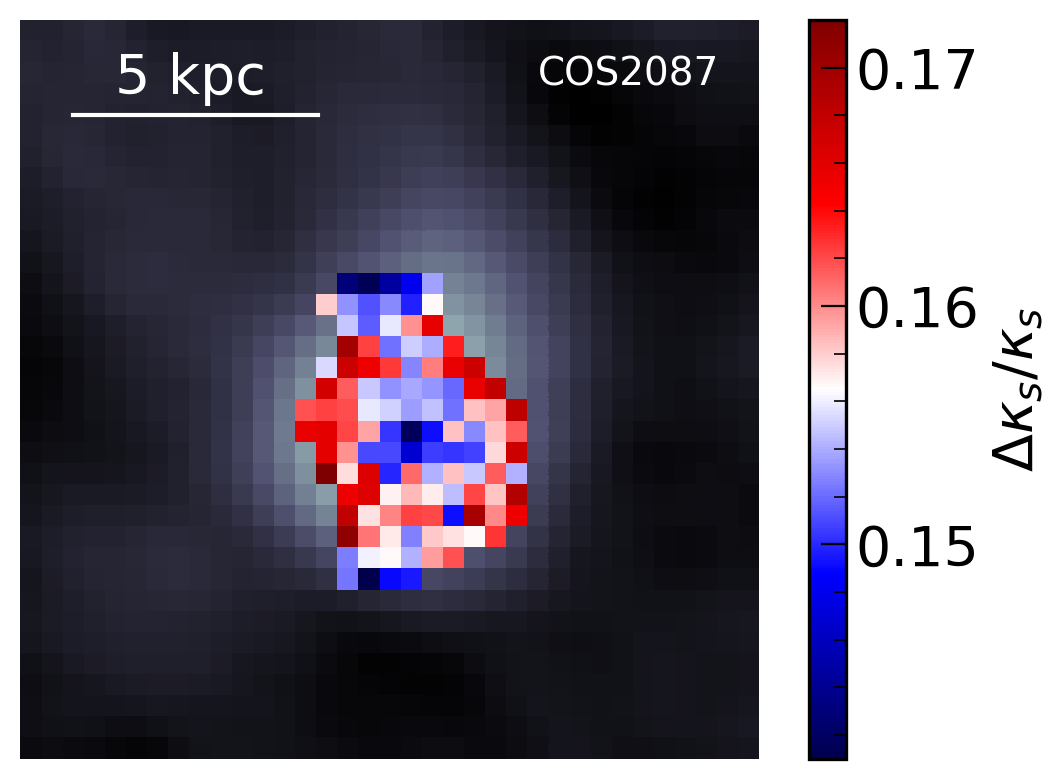}
    \includegraphics[scale=0.4]{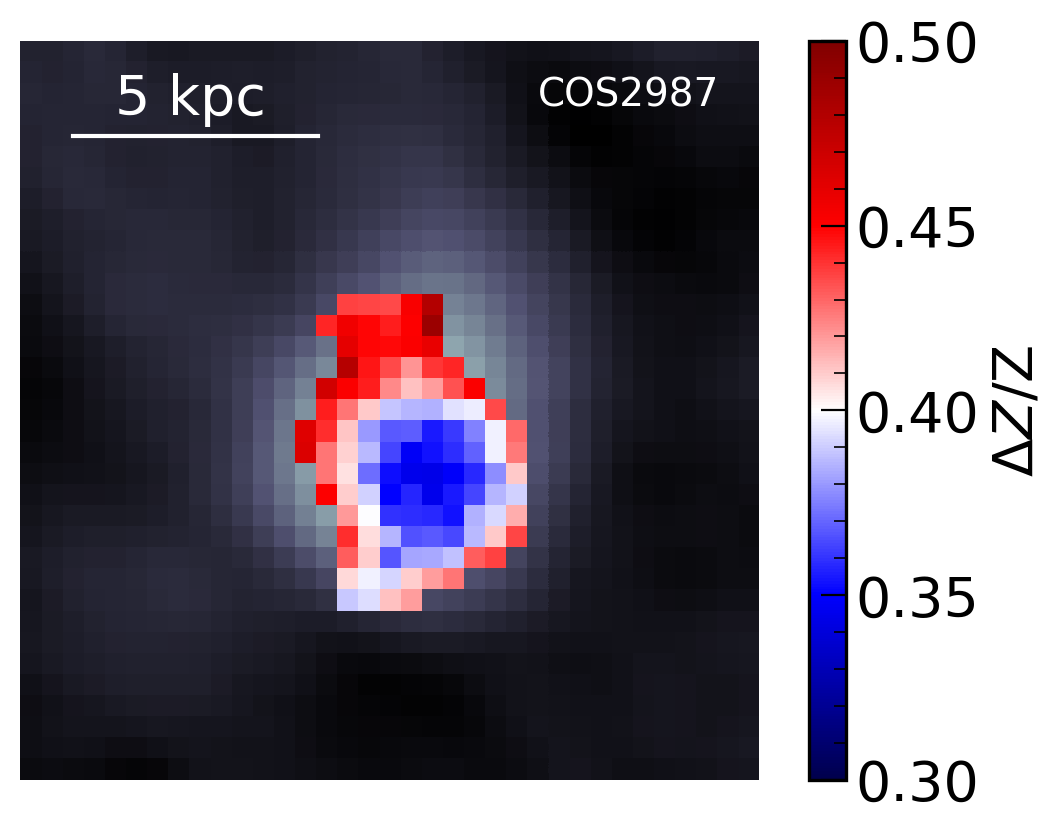}
    \includegraphics[scale=0.4]{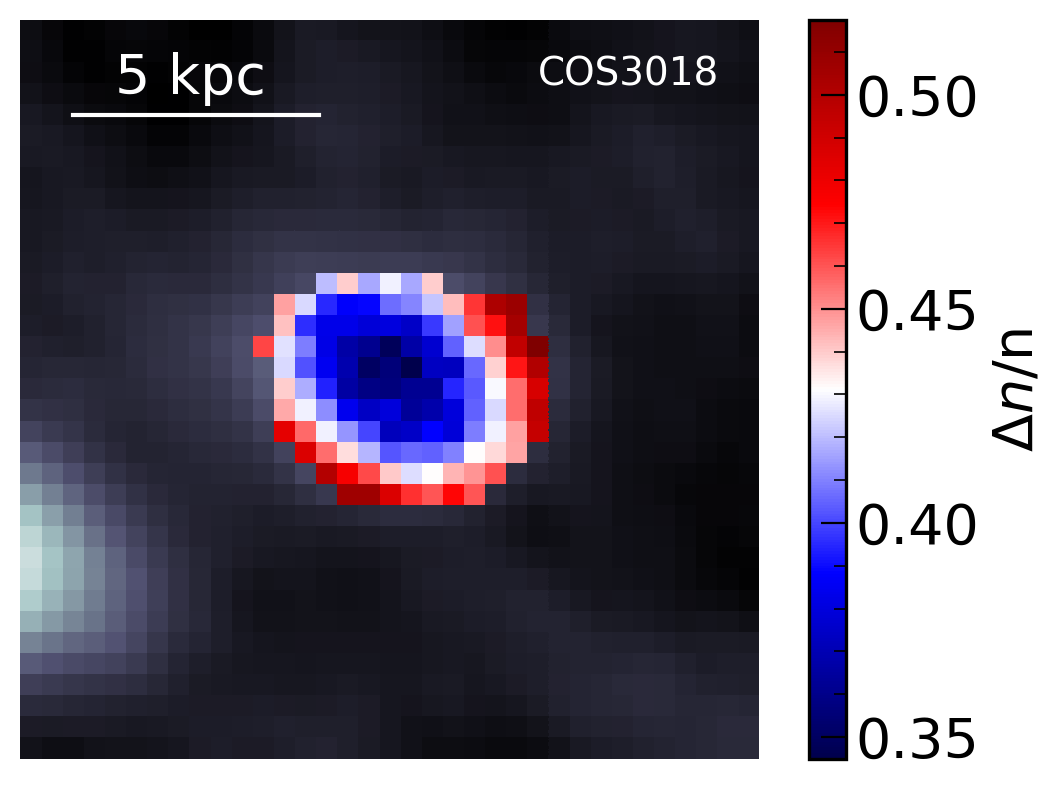}
    \includegraphics[scale=0.4]{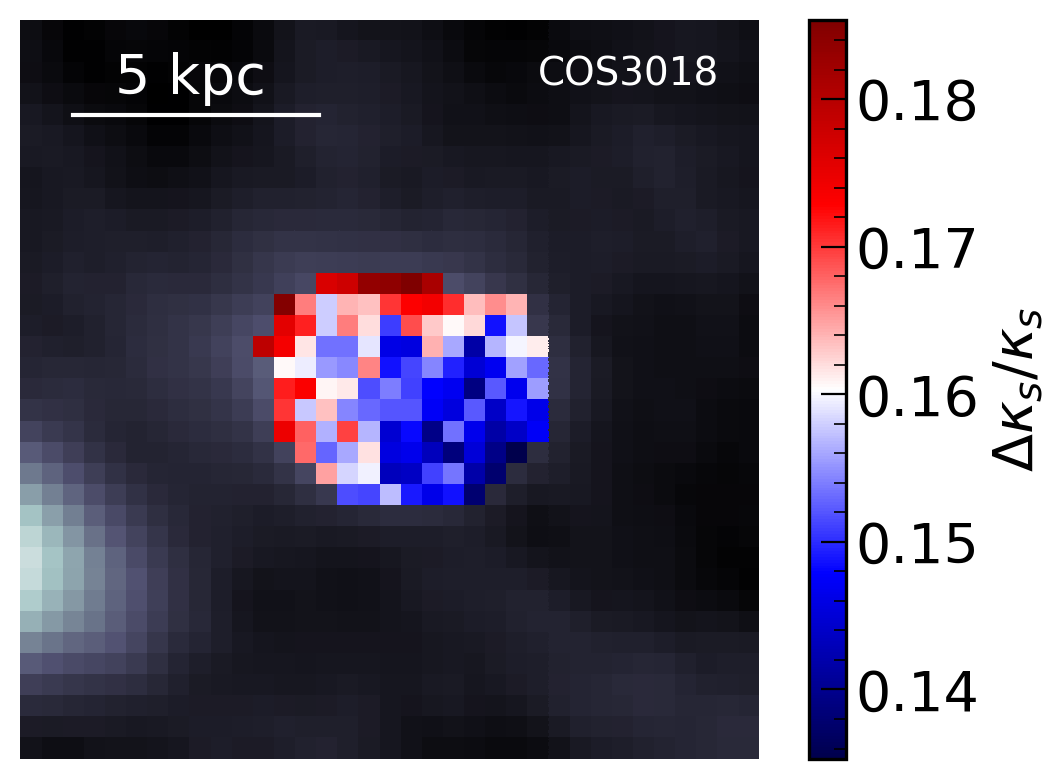}
    \includegraphics[scale=0.4]{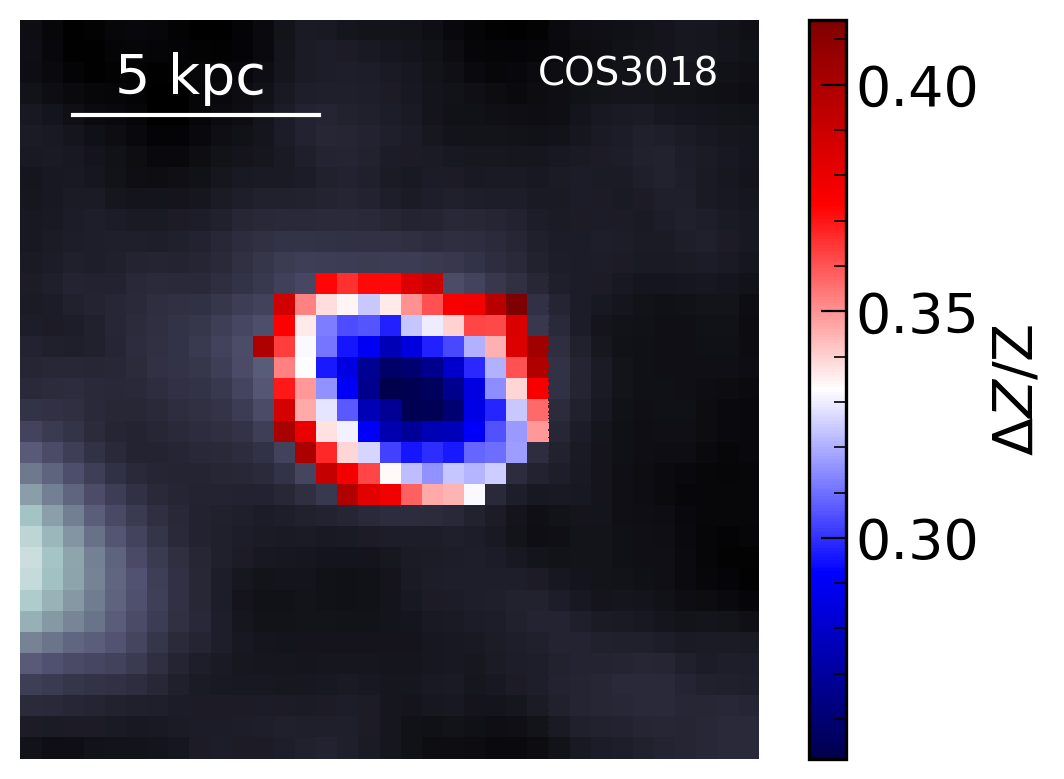}
    \includegraphics[scale=0.4]{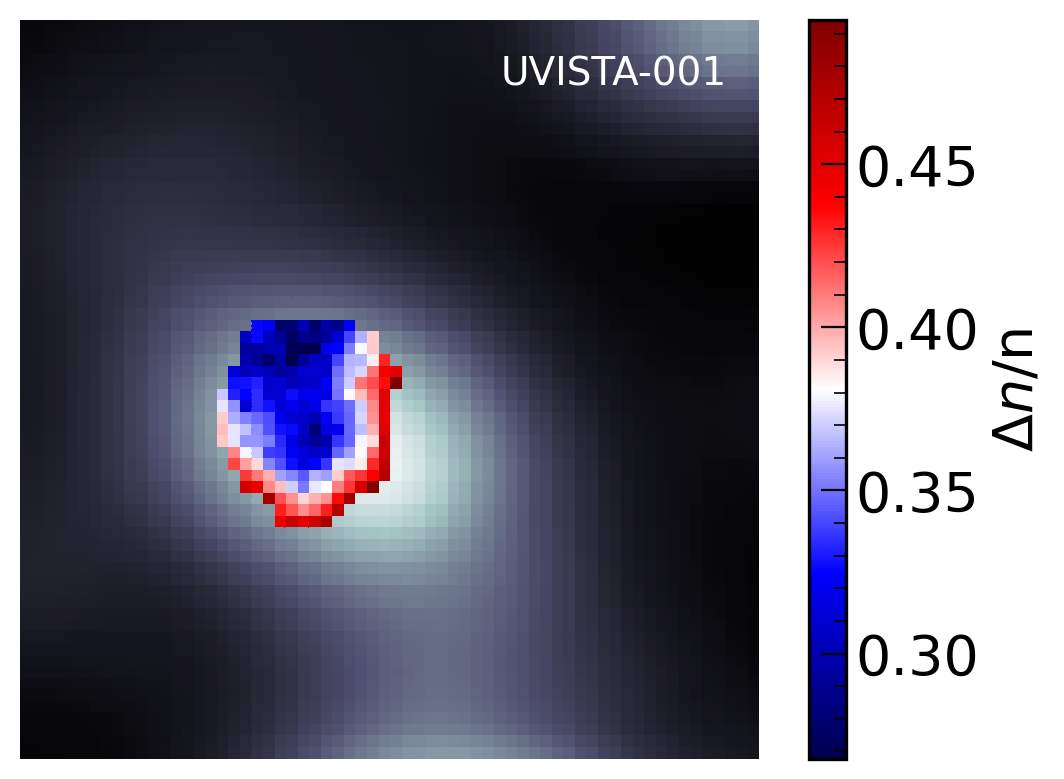}
    \includegraphics[scale=0.4]{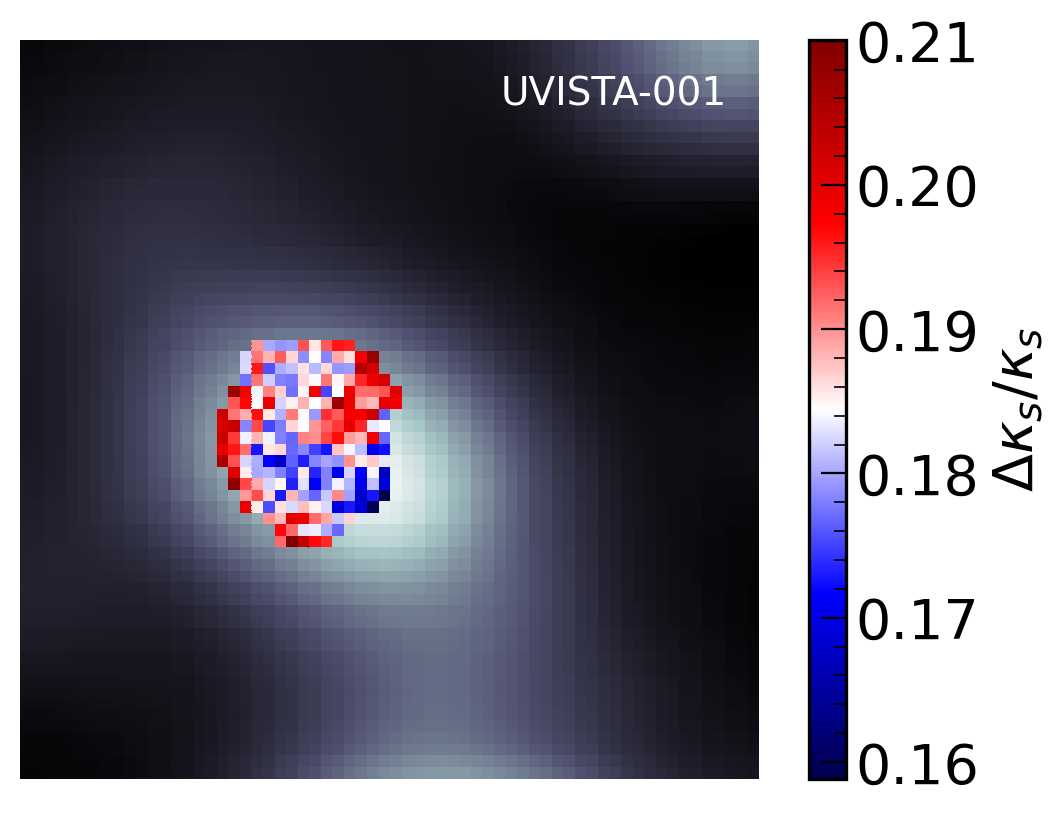}
    \includegraphics[scale=0.4]{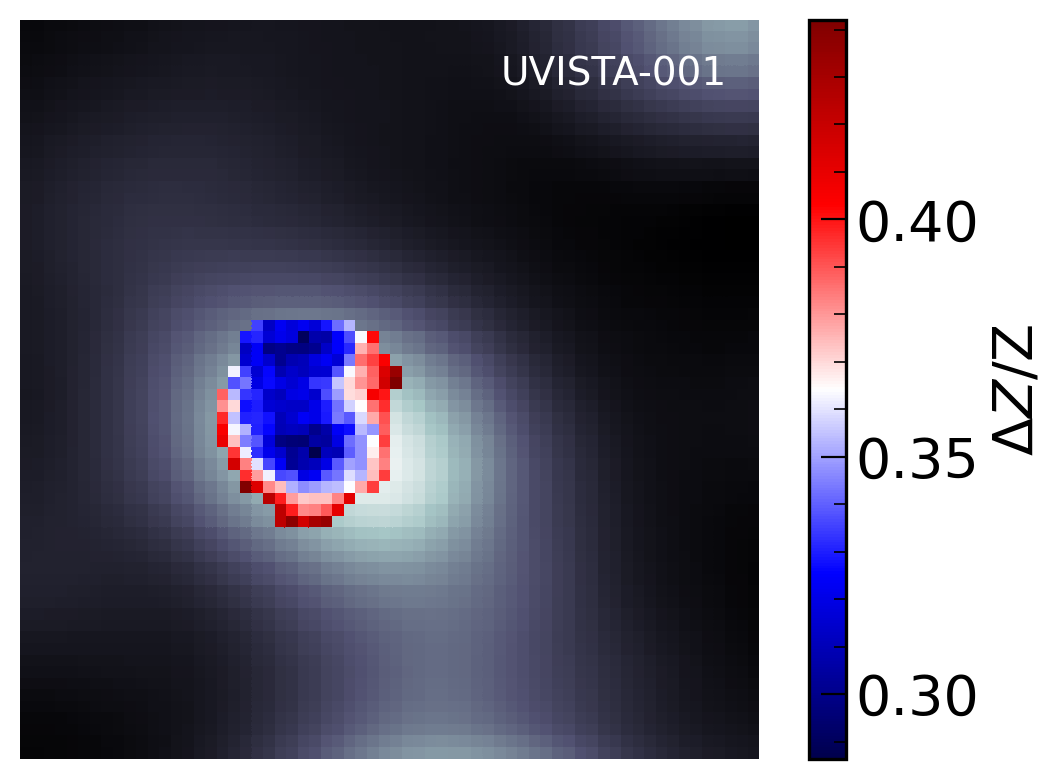}
    \includegraphics[scale=0.4]{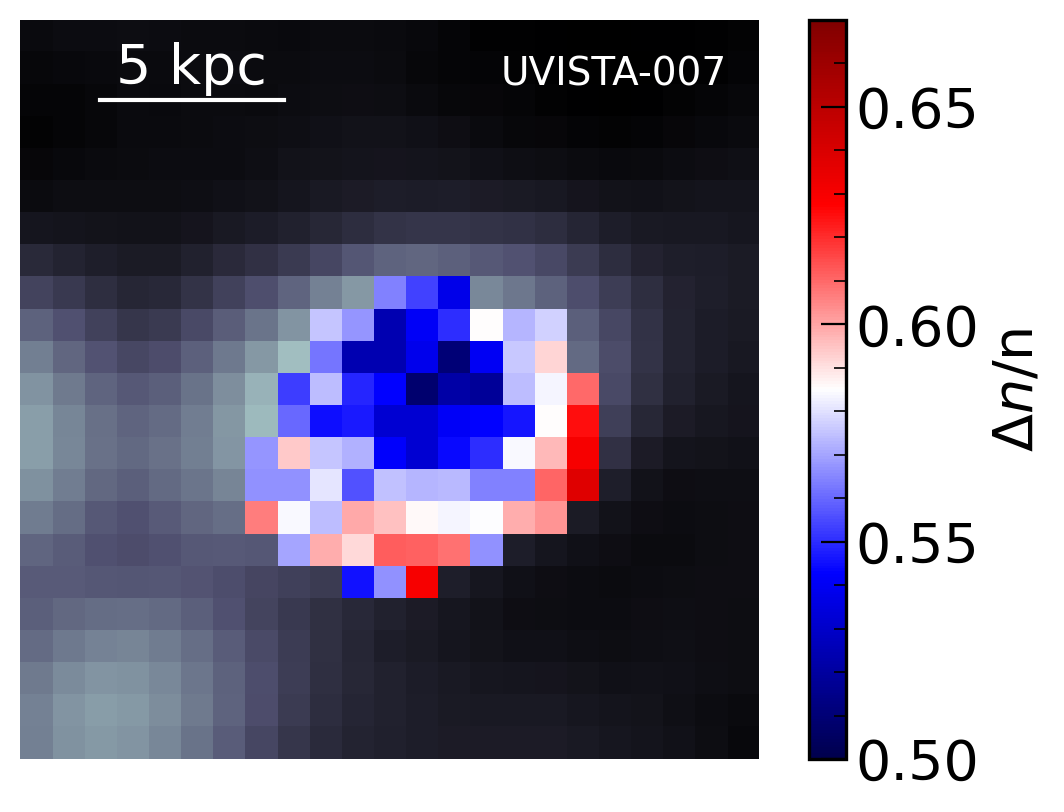}
    \includegraphics[scale=0.41]{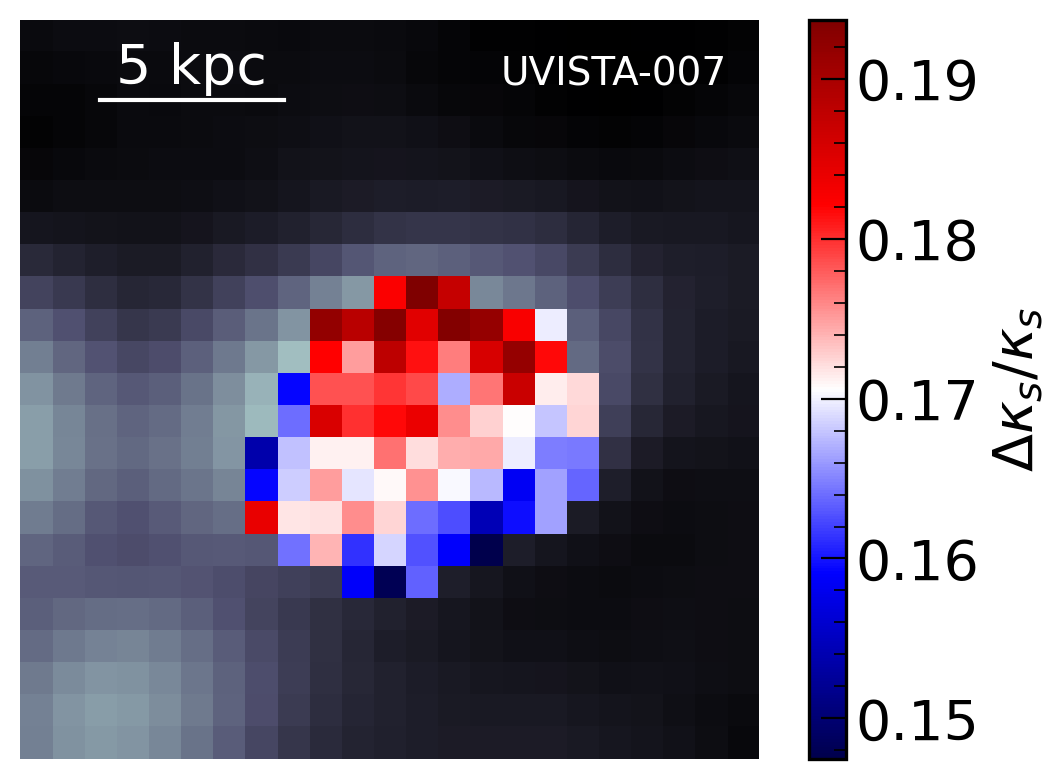}
    \includegraphics[scale=0.4]{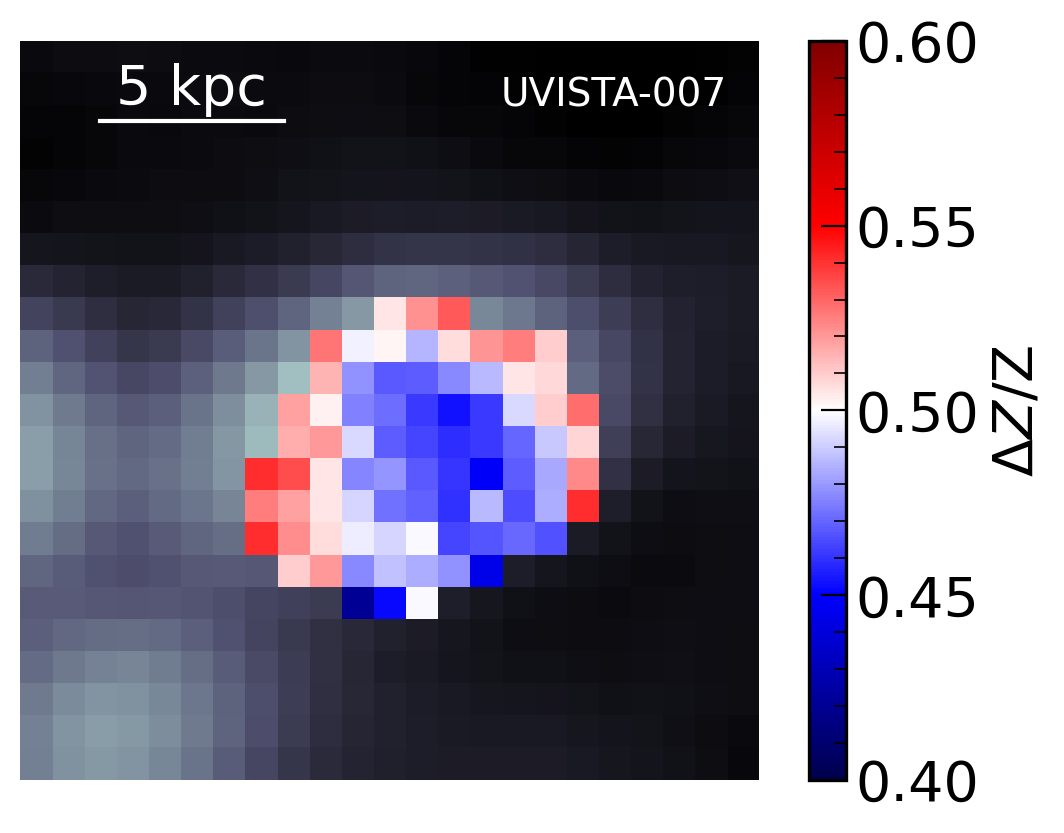}
    \includegraphics[scale=0.4]{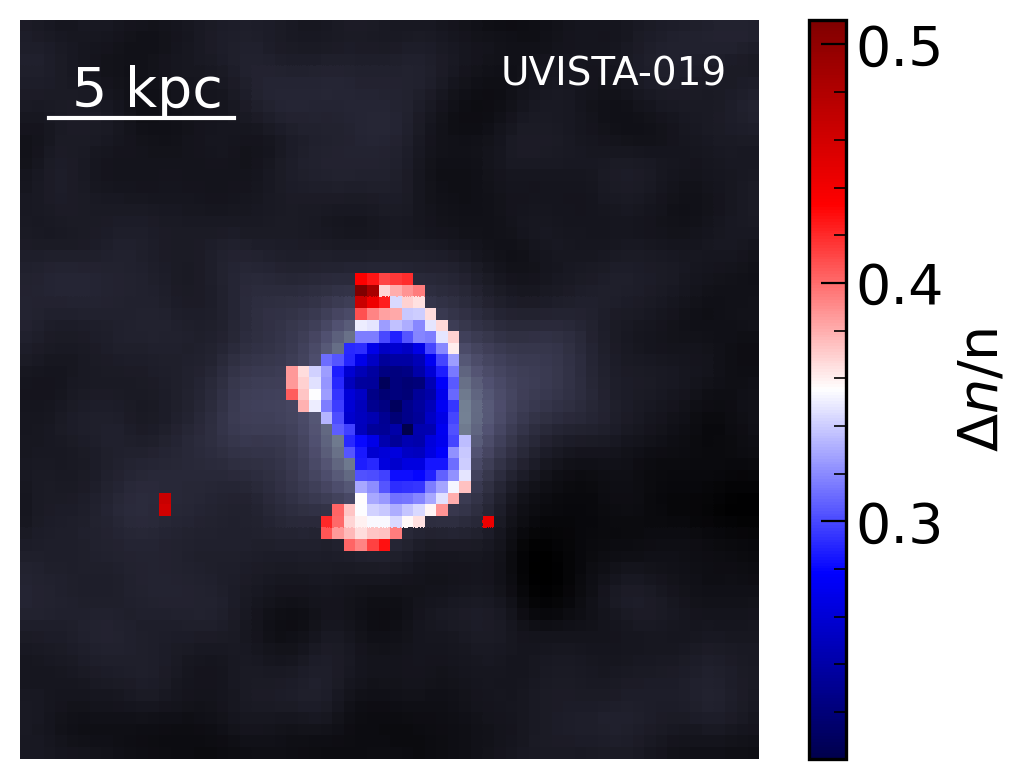}
    \includegraphics[scale=0.4]{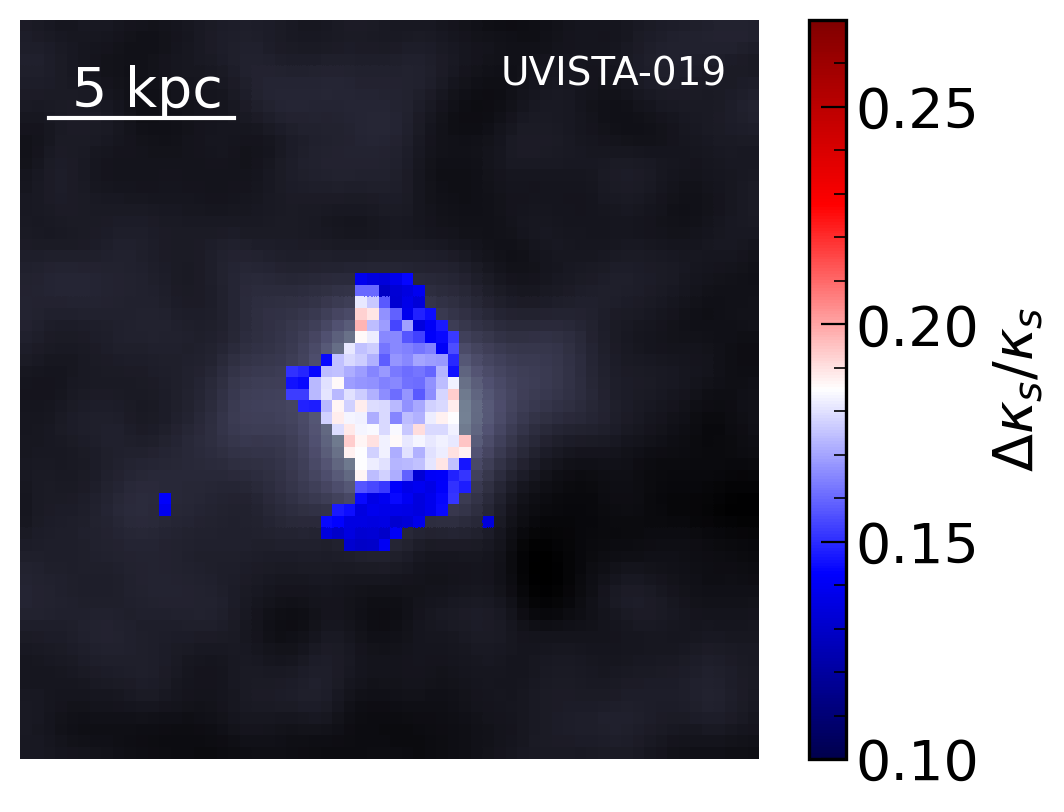}
    \includegraphics[scale=0.4]{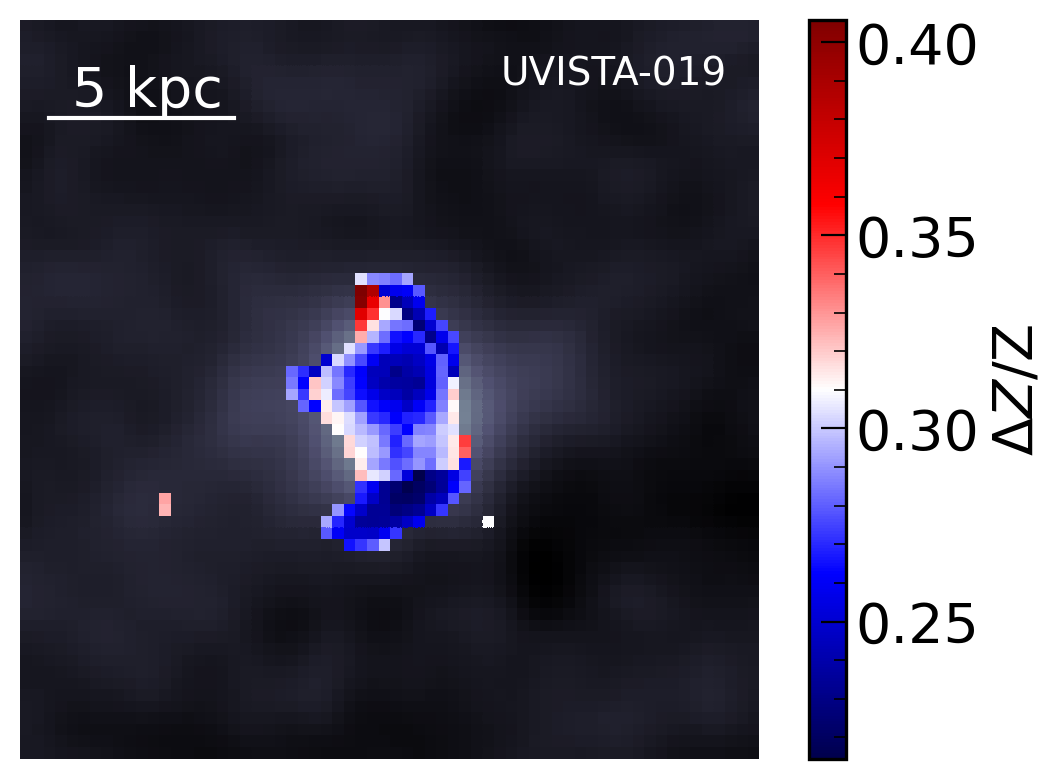}
    \caption{Maps of the relative errors on the gas density (left column), burstiness parameter (central column), and metallicity (right column) as obtained from \texttt{GLAM} on pixel by pixel basis.}
    \label{fig:GLAM_err}
\end{figure*}

In Figure \ref{fig:GLAM_err} we report the relative errors on the $(n, \, \kappa_s, Z)$ parameters derived with \texttt{GLAM} on pixel by pixel basis. The error on gas density ranges between 30\% -- 50\% depending on the source. The burstiness parameter is the ISM property that is better constrained, with errors ranging between 15\%--20\%. Finally, the metallicity errors range between 25\% -- 50\%.
{\color{black} We note that while the errors for the gas density and metallicity are lower in the central regions for all the galaxies, the trend is the opposite for $\kappa_s$ in UVISTA-Z-007 and UVISTA-Z-019, where the parameter is less constrained in the center. The uncertainties on ($n$, $\kappa_s$, $Z$) are influenced by the signal to noise ratio (SNR) of the inputs ($\Sigma_{\rm [CII]}$, $\Sigma_{\rm [OIII]}$ and $\Sigma_{\rm SFR}$),  which is higher in the center where the emission is brighter. However, this is not the only aspect influencing the uncertainty on the parameters derived from the MCMC. In fact, the analytical functions describing the \CII~and \OIII~fluxes are both characterized by a plateau in the flux at large $\Sigma_{\rm SFR}$ (see \citetalias{ferrara2019}), which makes the model more degenerate. For this reason there are regimes in which, the central part of the sources being characterized by large $\Sigma_{\rm SFR}$, the GLAM parameters end up being less precisely constrained albeit the SNR of the input data is higher. Improving the precision on the SFR tracers, with better rest-frame UV and IR data, can help in alleviating this issue.}
\section{Density and burstiness gradients}
\label{appendix:app_a}
\begin{figure}
    \centering
    \includegraphics[scale=0.45]{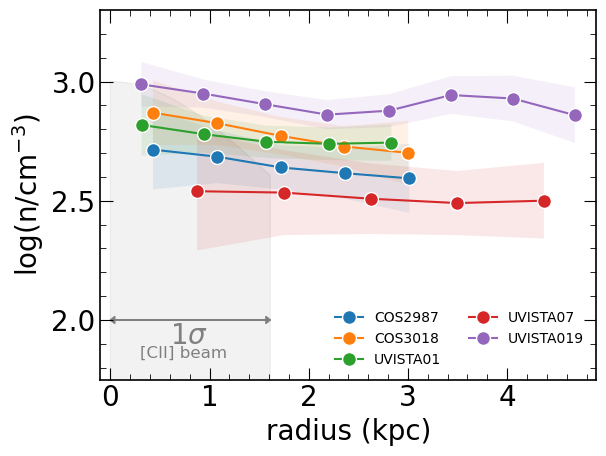}
    \includegraphics[scale=0.45]{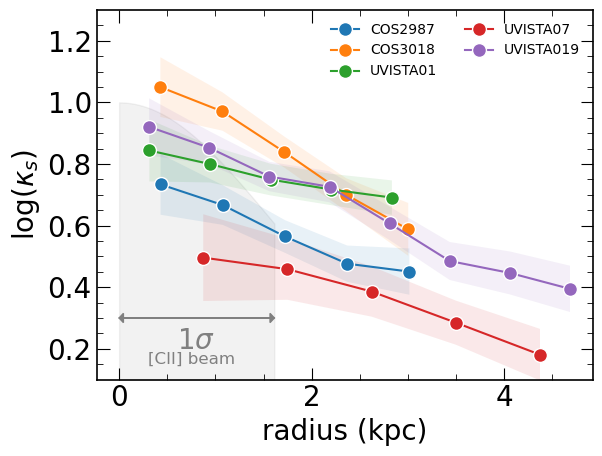}
    \caption{Radial profiles of the gas density ($n$, upper panel), deviation from the KS relation ($\kappa_s$, lower panel). The gray shaded region denotes the 1$\sigma$ width of the median \CII~beam. The shaded colored regions represent the $1\sigma$ error, see Sec \ref{sec:results} for details on the calculation.}
    \label{fig:app_a}
\end{figure}

We computed the gradient for the gas density and the burstiness parameter in the same way discussed for the metallicity ones. The results for our five sources are shown in Figure \ref{fig:app_a}. We note that, while the density profile is consistent with being flat (the median gradient among the five galaxies in our sample is $\nabla (\log n)\approx -0.01\pm 0.02$ dex/kpc), the burstiness parameter shows a steeper decrease (median $\nabla (\log \kappa_s)\approx -0.12 \pm 0.03$ dex/kpc). 


\bsp	
\label{lastpage}
\end{document}